%% file: main.tex
\documentclass[a4paper, 12pt, nofootinbib, amsmath, amssymb, aps, prd]{revtex4-2}

\usepackage{setspace}
\usepackage{CJKutf8} 
\usepackage{mathbbol} 
\usepackage{amssymb}

\usepackage[dvipsnames]{xcolor} 
\usepackage{hyperref} 
\hypersetup{
 colorlinks=true, 
 linkcolor=blue, 
 citecolor=PineGreen, 
 filecolor=magenta, 
 urlcolor=purple, 
 anchorcolor=blue,
 menucolor=Fuchsia,
 runcolor=magenta
}

\input{preamble}

\begin{document}
\begin{CJK*}{UTF8}{gbsn} 

\title{Quadratic gravity with propagating torsion and asymptotic freedom}
\author{Oleg Melichev}
\email{melichev@shanghaitech.edu.cn}
\affiliation{School of Physical Science and Technology, ShanghaiTech University 上海科技大学, 
\\ 393 Middle Huaxia Road, Pudong, Shanghai, 201210, China}

\begin{abstract}
We consider a class of metric-affine gravitational theories with action quadratic in curvature and torsion tensors.
Using the heat kernel technique, we compute the torsion contributions to the one-loop  
counterterms in the ultraviolet limit.
It is found that vectorial and axial components of torsion preserve the qualitative picture of the renormalization group flow of the metric sector.
However, there exists a specific nonminimal kinetic term for the pure tensorial (hook-antisymmetric traceless) component of torsion that renders the gravitational couplings asymptotically free in the absence of tachyons.
\end{abstract}

\maketitle


\section{\label{sec:intro}Introduction and Summary}

General Relativity (GR) can be embedded in the framework of quantum field theory, but this brings along several conceptual difficulties.
One of them is perturbative nonrenormalizability \cite{tHooft:1974toh,
Goroff:1985th,
Goroff:1985sz, 
vandeVen:1991gw}.
This fact, as well as some phenomenological issues that GR currently faces, prompted vast research in modified gravitational theories.
The most straightforward way to modify the gravitational equations of motion is to include higher curvature invariants into the action \cite{Stelle:1976gc, Modesto:2017hzl, Rachwal:2021bgb}.
Another way is to consider the affine connection as an independent variable \cite{Palatini:1919ffw}.
This yields two additional tensorial characteristics of the manifold, on top of the curvature: torsion, the antisymmetric part of the affine connection \cite{Einstein:1929b}, and the nonmetricity, the (minus) covariant derivative of the metric.
Metric-affine gravity (MAG) is a generic term for all theories that have at least one of them nonvanishing.
In this paper, we will assume that the nonmetricity vanishes, but torsion, on the opposite, is allowed to propagate and enter loops.
As we will see, this may have the potential to change certain renormalization group (RG) properties of the theory.

We aim to study the RG flow trajectories, primarily focusing on the asymptotically free ones.
The first such computations, performed for simplest Lagrangians with propagating torsion and nonmetricity, showed that, despite containing fourth-order terms in derivatives, they lead to a nonrenormalizable theory even at one-loop 
\cite{Melichev:2023lwj, Melichev:2024hih}.
For this reason, here we will instead consider MAG theories that are generalizations of metric Quadratic Four-Derivative Gravity (4DG).
This makes the power-counting renormalizability of the theory under consideration evident.
It is also well-known that the marginal couplings of 4DG are asymptotically free
\cite{Stelle:1977ry}.
The latter statement is however in tension with the condition of the absence of tachyons \cite{Salvio:2014soa}.
Tachyons are negative mass states that are definitely unhealthy even at low energies.
It has been shown recently that this tension can be circumvented when the condition for the asymptotic freedom is defined in terms of the momenta running \cite{Buccio:2024hys, Buccio:2024omv}.
In this paper, we will consider the $\m$-running of the Wilsonian approach \cite{Wilson:1973jj, Wilson:1974mb}, and show that in the case of propagating torsion, there might exist an even bigger room for asymptotically free theories without tachyons.

Due to the presence of rank-2 and rank-3 tensor fields, metric-affine gravity, as well as the usual metric gravity, is in general plagued with ghosts, which are solutions with negative kinetic energy.
Assuming certain relations between the couplings can ensure ghosts do not enter the spectrum \cite{Sezgin:1979zf, Sezgin:1981xs, Percacci:2020ddy, Lin:2018awc, 
Mikura:2023ruz,
Mikura:2024mji}.
However, it remains unclear whether quantum corrections will spoil the stability.
One can instead impose additional symmetries \cite{Marzo:2021iok}, but there exists an expectation that this program may also be in tension with renormalizability \cite{Marzo:2024pyn}.
Another approach to circumvent the ghosts and preserve unitarity was developed for metric 4DG in \cite{Anselmi:2018kgz, Anselmi:2018ibi} and further considered in the context of metric-affine gravity in \cite{Anselmi:2020opi, Piva:2021nyj, Piva:2023noi}.
It consists of changing the quantization prescription for the propagator from the usual Feynman one to the ``fakeon" prescription.
In the latter case, the ``ghosty" degrees of freedom are turned into purely virtual particles, which can appear inside loops but never as asymptotic states.
The theory then remains unitary \cite{Anselmi:2018kgz}, however, there remains an ambiguity of whether the healthy particles on top of the graviton are treated as physical or purely virtual.
It is important to note that such a ``fakeon" prescription only concerns the continuation into the Minkowski space and is not expected to generate any additional divergences.
Therefore, the expressions for the beta functions, which are commonly computed in the Euclidean space, are agnostic to whether we treat the ghost particles as purely virtual.
In this work, we focus on the RG flow, and we assume that all unhealthy degrees of freedom are either quantized as purely virtual or are harmless and do not spoil unitarity for another reason that is yet to be understood.
 
We want to briefly explain the origins of the main technical difficulty in calculating the quantum effective action in MAG theories, which comes from the kinetic operator structure.\footnote{Roughly speaking, one can think of the kinetic operator as the Hessian, \ie second variation of the classical action.
}
A kinetic operator of Laplace type is called \emph{minimal} if it only contains derivatives that are contracted with each other by their Lorentz indices, and not with fields.
Otherwise, it is called \emph{nonminimal}.
For example, an operator
\be \label{nonminopdef}
- \Box + V^\m \na_\m + E \, ,
\ee
where $\Box=g^{\m\n}\na_\m \na_\n$ and both $V$ and $E$ are free of derivatives, is minimal iff $V=0$.
In such cases, the beta functions can be obtained by directly applying the heat kernel technique \cite{Schwinger:1951nm, DeWitt:1964mxt}, for review see \cite{Vassilevich:2003xt}.
The application of this technique complicates significantly in the presence of nonminimal Laplace-type operators.
However, operators such as \eqref{nonminopdef} can still be easily dealt with by utilizing the fact that $V \ll \Box$ for high energy modes as $V$ is constructed of background quantities, and expanding the $\Tr\Log\!$ of the effective action in $V$.\footnote{Alternatively, one can define a new covariant derivative $\Tilde{\na
}_\m = \na_\m + 1/2\, V_\m$ and reabsorb the linear term into the quadratic one.}
Generically speaking, it is of crucial importance whether the \emph{principal part} of the kinetic operator (= its highest order in derivative terms) is minimal or nonminimal.
In MAG, we do face the latter situation, except for a measure-zero subspace of the theory space.
This means that we will have to deal with kinetic operators such as $\na^\m \na_\n$ and $\na_\m \Box$ inside its principal part, and most importantly, such terms cannot be eliminated by choice of gauge, except for some specific Lagrangians that have been studied in \cite{Melichev:2023lwj, Melichev:2024hih}.
The possible ways of dealing with them include the transverse-longitudinal decomposition and the ``off-diagonal" heat kernel technique.
The former method is analogous to York decomposition \cite{York:1973ia} in metric theories, and is most convenient for calculations on maximally symmetric backgrounds.
In general, operators can also be transformed into the minimal form with specific choices of the background, but this hinders the ability to ``read off" the beta function of the Weyl squared term.
See, for example, \cite{Ohta:2016jvw, Ohta:2016npm} for such computations in Einstein theory and quadratic gravity.
The calculation of \cite{Martini:2023apm} also follows this strategy.
It was built up in relation to torsion theories in \cite{Martini:2023rnv, Sauro:2024lyl} and applied for the computation of contributions to the running of the Hilbert--Einstein and Starobinsky $R^2$ terms from the fluctuations of torsion in \cite{Martini:2023apm}.
We will use instead the ``off-diagonal" heat kernel technique \cite{Groh:2011dw}, also known as the
``generalized Schwinger--DeWitt technique", as originally introduced by A.~Barvinsky and G.~Vilkovisky in \cite{Barvinsky:1985an}.
It is most convenient for operators with nondegenerate principal part (which relates to the ``condition of causality" described in \cite{Barvinsky:1985an}).
This will be commented on in the next section.
The ``off-diagonal" heat kernel method is also applied in the context of non-perturbative calculation with the functional renormalization group \cite{Benedetti:2010nr, Knorr:2021slg}.

We will examine the case when the action contains kinetic terms for torsion, causing torsional degrees of freedom, as well as the metric ones, to enter the loops. 
Since torsion has the mass dimension one, they naturally come as operators of dimension four.
A major difficulty about such operators is that they often have nonminimal structure.
In general, at order four in mass dimension, metric-torsion gravity has 3 minimal and 6 nonminimal contributions of the type $\left( \nabla T \right)^2$, as well as two mixing terms of the type $R\nabla T$ and three quadratic in curvature invariants.
All these terms should be taken into consideration, unless they are prohibited by symmetry or can be reabsorbed into field redefinitions.

Gravitational theories with propagating torsion have received significant attention 
\cite{Neville:1979rb,
Hammond:1990aw, 
Saa:1996mt, 
2004.14776, 
Baikov:1992uh, 
Tseytlin:1981nu, 
Hernaski:2009wp, 
Sezgin:1981xs, 
Deveras:2010wq}.
There have also been attempts to compute the UV divergences \cite{Tseytlin:1981nu, Piva:2021nyj, Piva:2023noi,
Percacci:2023rbo}, and we will further generalize some of these results.
Asymptotically safe Hilbert--Palatini gravity was studied in an on-shell reduction scheme \cite{Gies:2022ikv}.
Non-perturbative beta functions for MAG with non-propagating torsion and nonmetricity were found in \cite{Pagani:2015ema, Reuter:2015rta}.

Torsion generically contains three components, which are irreducible under the Lorentz group: vectorial, axial, and pure tensorial (hook-antisymmetric traceless). 
In this paper, we compute the separate one-loop contributions of these three components to the running of the gravitational couplings in front of quadratic curvature invariants.
As expected, all minimal terms, as well as the nonminimal terms of the vectorial and axial sectors, do not produce sufficient contributions to change the sign of the beta functions.
On the other hand, it was found that one of the nonminimal terms of the kinetic sector for the hook-antisymmetric component can contribute enough to make the beta function of $R^2$ change the sign while simultaneously preserving the negative sign in front of the $\text{Weyl}^2$ term, which renders the corresponding couplings asymptotically free in the absence of tachyons.

The rest of this paper is organized as follows.
We formulate the model at consideration in section \ref{sec:kinetic}.
We elucidate the application of the heat kernel technique in the context of nonminimal operators, which can also find various applications outside MAG, in section \ref{sec:divergences}.
We obtain the results for the beta functions in section \ref{sec:betas} and proceed with their discussion in section \ref{sec:discussion}.
The general form of the results, together with some intermediate formulae, are moved to the appendices  \ref{sec:app.results} and \ref{sec:app.relations}.
We use the mostly positive signature.

\section{\label{sec:kinetic}Kinetic structure of gravity with propagating torsion}

\subsection{\label{subsec:Lags}Lagrangian}

The most general parity-even action that is quadratic in curvature and torsion reads \cite{Christensen:1979ue, Diakonov:2011fs, Baldazzi:2021kaf}:
\bear \label{action}
S & = - \polov \int d^4 x \sqrt{g} \Big[ 
m_p^2 \left( 2\L - R \right)
+ 2 \a R^2 
+ 2 \b R_{\m\n} R^{\m\n} 
+ 2 \g R_{\m\n\r\s} R^{\m\n\r\s}
\\ & 
- b_1\, T_{\m\n\r} \Box T^{\m\n\r} 
- b_2\, T_{\m\n\r} \Box T^{\m\r\n} 
- b_3\, \tv_{\m} \Box \tv^{\m} 
- b_4\, T^{\m\r\l} \na_{\m} \na_{\n} T^{\n}{}_{\r\l} 
\\ &
- b_5\, T^{\m\r\l} \na_{\m} \na_{\n} T^{\n}{}_{\l\r} 
- b_6\, T^{\r\m\l} \na_{\m} \na^{\n} T_{\l\n\r}
- b_7\, T^{\r\m\l} \na_{\m} \na^{\n} T_{\n\r\l} 
\\ &
- b_8\, T^{\r\m\l} \na_{\m} \na^{\l} \tv^{\r}
- b_9\, \tv^{\m} \na_{\m} \na_{\r} \tv^{\r} 
+ \t R \na_{\m} \tv^{\mu} 
+ \s R^{\m\n} \na_{\r} T^\r{}_{\m\n} 
\\ &
- m_1\, T_{\m\n\r} T^{\m\n\r} 
- m_2\, T_{\m\n\r} T^{\m\r\n} 
- m_3\, \tv_{\m} \tv^{\m} 
+ \dots
\Big] \, ,
\eear
where we defined the torsion vector $\tv_\m = T_\m{}^\a{}_\a$, $R_{\m\n\r\s}$ is the Riemann curvature of the Levi-Civita connection, and $\sqrt{g} = det |g_{\m\n}|$.
The terms of the first row correspond to the usual quadratic gravity action \cite{Stelle:1976gc, Salvio:2018crh}.
The next three rows contain nine terms of the kinetic matrix for the torsion and two possible mixing terms with curvature and torsion.
We can see that the first three of them, $b_1$, $b_2$, $b_3$, are minimal while the other six $b_4$, \dots , $b_9$ are nonminimal.
The mixing terms $R\na T$ also produce nonminimal operators after perturbative expansion.
The terms of the last row introduce masses for the torsional degrees of freedom.
The ellipses refer to the fact that one could also include other terms, such as $T^4$, $RT^2$, and $T^2\na T$ that do not contribute to the 2-point function around flat space, as well as terms of higher mass dimension, such as $R\Box R$ or $T\Box^2 T$.
In this paper, we will disregard such contributions and focus on the expansion up to quadratic order.

The action \eqref{action} is written in what we call the \emph{``Einstein formulation"}, which means using Riemann curvatures and Levi-Civita derivatives.
In other words, the derivatives $\na$ have zero torsion, while the curvatures are functions of the metric only.
One could instead choose to write it via Cartan curvatures and the independent torsion-full connection.
Although geometrically well-motivated, such \emph{``Cartan formulation"} in practice leads to longer computations and often mistakes.
The relation between these formulations has been extensively discussed in \cite{Baldazzi:2021kaf}.
It is always possible to convert the results obtained in one formulation into the other if one so desires.

Before discussing the computation of divergences in detail, it is illustrative to look at the general form of the kinetic operator, obtained by the second variation of the action.
What we will need for the one-loop computation is the trace of the logarithm of that operator, which, roughly speaking, can be obtained if we know its inverse.
The complexity of the heat kernel technique application for a particular computation hinges upon the structure of the principal part of the kinetic operator.
Very schematically, the kinetic operator and its inverse have the following forms:
\be
\label{kin_op_der_structure}
S^{(2)} = \begin{pmatrix}
\Box^2
& \Box \na
\\
\Box \na & \na\na 
\end{pmatrix} + \mathcal{O} \left(\bar{R},\bar{T}\right) \, ,
\quad\quad\quad
G = \frac{1}{\Box^4} \begin{pmatrix}
\na^4
& \Box \na^3
\\
\Box \na^3 & \Box^2 \na^2
\end{pmatrix} + \mathcal{O} \left(\bar{R},\bar{T}\right) \, .
\ee
Here we visualize this matrix acting on the sets of fields $\varphi = (h_{\m\n}, \d T_\r{}^\s{}_\l)^T$.
The Hessian matrix $S^{(2)}$ contains metric perturbations $h_{\m\n}$ coming with fourth, torsion perturbations with second, and mixing terms with third-order operators.
It is therefore crucial to check whether the unique inverse of this form exists.
In such cases, we will say that the kinetic operator has \emph{nondegenerate principal part}.
If that is not true \emph{even after the gauge is fixed}, we will say that the principal part is \emph{degenerate}.
As the simplest example of the degenerate case, one can think of the Proca theory.
The kinetic operator cannot be inverted directly, because the longitudinal component enters only the mass term.
To solve the issue and eventually compute the divergences, one can employ the St\"uckelberg trick.
Similar situations can happen in theories with propagating torsion and nonmetricity for specific choices of the couplings of the Lagrangian.
The first study of the degenerate case with a healthy propagating vector was performed in \cite{Marzo:2024pyn}, revealing several possible inconsistencies of this approach.
In this paper, we look only at \emph{nondegenerate} cases, where the problem described above does not arise.
This choice highlights the issue already described in the previous section.
Most choices of the Lagrangian, especially the ones that lead to kinetic operators with nondegenerate principal parts, produce unphysical particles with the wrong sign of kinetic energy.
For this reason, we must assume that these particles appear only inside loops, and never as asymptotic states. 

The computational method that we use is perfectly applicable in the most general setup of \eqref{action}.
Moreover, it allows the computation of all the counterterms on backgrounds with arbitrary torsion and curvature.
However, due to the complicated kinetic mixing that it leads to, the final answer for the beta functions will be a cumbersome rational function of the couplings, which at first can be hard to make sense of.
For this reason, we will make two simplifications here.
Firstly, we will only consider torsion-free backgrounds.
This is sufficient in order to see the running of the couplings $\a$, $\b$, and $\g$ (or alternatively, $\l$, $\x$, and $\r$ when the action is written in the Weyl basis, see below), which is our purpose.
Secondly, we will consider the contributions of different components of torsion separately.

Indeed, the torsion tensor can be decomposed into irreducible representations of the Lorentz group as
\be \label{torsion_decomposition_def}
\tv_{\m} = T_{\m}{}^{\a}{}_{\a} \, , 
\quad\quad
\ta_{\m} = \epsilon_{\m\n\r\l} T^{\n\r\l} \, , 
\quad\quad
\tp_{\a\b\g} = T_{\a\b\g} - T_{[\a\b\g]} - \frac{1}{6} \tv_{[\a} g_{\b|\g]} \, , 
\ee
where $\tv_\m$ is a vector, $\ta_\m$ is an axial vector and $\tp$ is the pure tensorial part of torsion that is traceless and hook-antisymmetric:
\be \label{tensorial_torsion_properties}
\tp_{\m}{}^{\a}{}_{\a}{} \equiv 0, \quad \epsilon^{\m\n\r\l} \tp_{\n\r\l} \equiv 0 \, ,
\ee
with $\epsilon_{\m\n\r\l}$ being the Levi-Civita symbol.
By substituting \eqref{torsion_decomposition_def} into \eqref{action} we obtain, up to quadratic order,
\bear \label{action_decomposed}
S &= - \polov \int d^4 x \sqrt{g} \Big[
m_p^2 \left( 2 \L - R \right)
+ \frac{1}{\l} C^2 
- \frac{1}{3\x} R^2
+ \frac{1}{\r} E_{\rm GB}
\\&
+ r_1\, R~ \na_{\m} T^{\m} 
+ r_2\, C_{\m\n\r\s} \na^{\m} \tp^{\n\r\s} 
- d_1\, T_{\m} \Box T^{\m} 
- d_2\, T_{\m} \na_{\m} \na^{\n} T^{\n}
\\&
- d_3\, \ta_{\m} \Box \ta^{\m} 
- d_4\, \ta_{\m} \na_{\m} \na^{\n} \ta^{\n} 
- d_5\, \tp_{\m\n\r} \Box \tp^{\m\n\r}
- d_6\, \tp_{\m\r\l} \na^{\m} \na_{\n} \tp^{\n\r\l} 
\\&
- d_7\, \tp_{\m\r\l} \na^{\m} \na_{\n} \tp^{\n\l\r}
- d_8\, T_{\r} \na_{\m} \na_{\n} \tp^{\r\m\n}
- d_9\, \lctens_{\m\n\r\l} T^{\m} \na^{\l} \na_{\s} \tp^{\s\n\r} 
\\&
+ m_{\tv}\, \tv_{\m} \tv^{\m} 
+ m_{\ta}^2\, \ta_{\m} \ta^{\m} 
+ m_{\tp}^2\, \tp_{\m\n\r} \tp^{\m\n\r}
+ \dots
\Big] \, .
\eear
Here $C^2 = C^{\m\n\r\s} C_{\m\n\r\s}$ is the square of the Weyl tensor, and $\lctens_{\m\n\r\l} = \sqrt{g}\; \epsilon_{\m\n\r\l}$ is the Levi-Civita tensor.
The combination $E_{\rm GB} = R_{\m\n\r\s}R^{\m\n\r\s} - 4 R_{\m\n}R^{\m\n} + R^2 $ relates to the Euler--Gau\ss--Bonnet invariant. 
Since it is constructed with Riemann curvature of the Levi-Civita connection, it does not contribute to the equations of motion, just as in Riemannian geometry.\footnote{See \cite{Babourova:1996hy, Janssen:2019uao} for related discussions.}
Going from \eqref{action} to \eqref{action_decomposed} involves the utilization of several relations that are consequences of \eqref{tensorial_torsion_properties}.
We describe this technical step in more detail in the App. \ref{sec:app:Lags}.

\subsection{\label{subsec:hessian}The Hessian structure and gauge fixing}

For a one-loop calculation within the background field method, the first step would be to compute the second variation of the action \eqref{action_decomposed} with respect to fields: 
\be \label{hess_def}
S^{(2)}_{ab} = \frac{\d^2 S}{\d \varphi^a \d \varphi^b} \, .
\ee
Here we adopted the condensed DeWitt notations for the field perturbations as a column $\varphi^{a}$, where index $a$ carries both all the internal structure and the space-time dependence:
\begin{equation}
\label{column_of_fields_in_MTG}
\varphi^{a} = \begin{pmatrix} h_{\mu\nu}
\;&\; \d \tv_\a 
\;&\; \d \ta_\b
\;&\; \d \tp_{\rho}{}^{\l}{}_{\eta} \end{pmatrix}^{T} (x).
\end{equation}
On the configuration space, one can define metric, torsion, and nonmetricity, which can be arbitrary in general.
Since we have chosen to work with the Levi--Civita derivative, the latter two vanish.
This fact allows the application of the standard heat kernel expansion.
We also utilize the simplest ultralocal configuration space metric of the form
\be
\label{configuration_space_metric}
\cG_{ab} = \begin{pmatrix}
\mathbb{1}_{\text{s}}^{\;\a\b ,\; \m\n}
\quad&\quad g^{\t\s} \quad&\quad g^{\omega\upsilon} \quad&\quad \mathbb{1}_{\text{hat}}{}^{\,\g}{}_{\d}{}^{\eta ,\; \r}{}_{\l}{}^{\z} 
\end{pmatrix} \delta (x - x') \, .
\ee
The tensors $\mathbb{1}_{s}$, $\mathbb{1}_{\text{hat}}$, and $\mathbb{1}_{\text{at}}$ are the identities in the spaces of symmetric rank-2, hook-antisymmetric traceless rank-3, and antisymmetric traceless rank-3 tensors respectively:
\begin{subequations}
\begin{align} 
& \label{id_h}
\mathbb{1}_{\text{s}}^{\;\a\b ,\; \m\n} = \polov \left( g^{\a\m} g^{\b\n} + g^{\a\n}g^{\b\m} \right) \, ,
\\& \label{id_hat}
\mathbb{1}_{\text{hat}}{}^{~\g}{}_{\d}{}^{\eta ,\; \r}{}_{\l}{}^{\z} = \mathbb{1}_{\text{at}}{}^{\,\g}{}_{\d}{}^{\eta ,\; \r}{}_{\l}{}^{\z} 
- 
\frac{1}{6} \epsilon^{\g}{}_{\r}{}^{\eta \eta_1} \epsilon_{\eta_1 \r_1}{}^{\l_1}{}_{\z_1} 
\mathbb{1}_{\text{at}}{}^{\,\r_1}{}_{\l_1}{}^{\z_1, \; \r}{}_{\l}{}^{\z} \, ,
\\& \label{id_at}
\mathbb{1}_{\text{at}}{}^{\,[\g}{}_{\d}{}^{\eta] ,\; [\r}{}_{\l}{}^{\z]} = g^{\g\r} g_{\d\l} g^{\eta\z} - \frac{2}{3} \d^{\g}_{\d} \d^\r_\l g^{\eta\z} \, .
\end{align}
\end{subequations}
The right-hand side of the last term is understood to be antisymmetrized in both $[\g \eta]$ and $[\r\z]$.

After computing the Hessian and with our choice of the configuration space metric, one obtains the expression for the kinetic operator.
Its full form is too long to be presented here.
However, it is important to study the part that is nonvanishing in the flat space limit ($R=T=0$).
Expectedly, it is not invertible due to gauge freedom.
We can use the following condition to fix the gauge:
\be \label{gauge_chi_def}
\chi_\m = \na^\l h_{\l\m} + b\, \na_\m h + c\, \d T_\m = 0 \, ,
\ee
which can be enforced by adding gauge fixing terms to the action in a way that is standard for metric 4DG \cite{deBerredo-Peixoto:2004cjk}:
\be \label{S_gf}
S_{\text{gf}} = \frac{1}{2 a} \int d^4 x \sqrt{g}\; \chi_\m C^{\m\n} \chi_\n \, ,
\ee
where the weight operator is
\be \label{3rd_gh_op_def}
C^{\m\n} = \bar{g}^{\m\n} \bar{\Box} + e \bar{\na}^\m \bar{\na}^\n + f \bar{\na}^\n \bar{\na}^\m + d\, m_p^2\, \bar{g}^{\m\n} \, .
\ee
The terms of the kinetic operator obtained after the gauge fixing that are nonzero in flat space are explicitly shown in \ref{subsec:kinop_expression}.
We can choose 
\be \label{Gauge_Choice_Percacci}
a = \l \, , \quad 
b = -\frac{2 \l+\x}{2 \l+4 \x}\, , \quad 
c = d = 0\, , \quad 
e = \frac{\l+2\x}{3\x}\, ,
\quad f = 1\, .
\ee
With this choice, the $hh$ sector of the kinetic operator becomes minimal.
In general, however, there is no such choice of gauge fixing condition that would eliminate all the nonminimal terms, so we will have to deal with them separately.
We will see how they contribute to the effective action in the next section.

\section{\label{sec:divergences}Computation of divergences}

Taking into account the contributions from the Faddeev--Popov ghosts and the third ghost operator \eqref{3rd_gh_op_def}, we have the following expression for the one-loop effective action \cite{Feynman:1963ax, DeWitt:1967yk, 
Faddeev:1967fc,
Buchbinder:1992rb,
Buchbinder:2017lnd, Percacci:2017fkn}:
\be \label{1-loop_EA_with_ghosts}
\G^{\rm 1-loop} = \polov \Tr\Log F - \Tr\Log \Delta_{gh.} - \polov \Tr\Log C_{3gh.} \, .
\ee
After the gauge fixing, the
part of the kinetic operator $F$ that does not vanish in flat space can be represented as
\be \label{X_Fmin_N_split}
X (\lamBar) = X_{\rm min} + \lamBar N \left( \na \right) \, ,
\ee
where the first term represents the minimal part
\bear \label{Fmin_expr}
X_{\rm min} &= \frac{\b+4\g}{2} {h}{}^{\m\n} \Box^2 h_{\m\n} 
+ \frac{(4\a+\b)(\b+4\g)}{8(\a-\g)} h{} \Box^2 h 
\\&
- d_1 {\d \tp}{}^\r \Box \d \tp_\r 
- d_3 {\d \ta}{}^\r \Box \d \ta_\r 
- d_5 {\d \tp}{}^{\m\n\r} \Box \d \tp_{\m\n\r} \, ,
\eear 
and $N ( \na)$ stands for all the nonminimal therms (those containing uncontracted derivatives).
The factor $\lamBar$ is introduced for later use, here $\lamBar=1$.
The full kinetic operator is then
\be \label{kinop_XY_split}
F (\lamBar) = X(\lamBar) + Y \!\left( \bar{R}, \bar{\tv}, \bar{\ta}, \bar{\tp} \right) \, .
\ee
In the computation performed in this paper, we assume a torsion-free background, but the same reasoning applies in the general case.
Then we use the following trick \cite{Barvinsky:1985an}:
\be \label{Barvinsky_trick}
\Tr\Log X (\lamBar=1) = \Tr\Log \left[ X_{\rm min} + \lamBar N(\na) \right]_{\lamBar=1} = \Tr\Log X_{\rm min} + \int_0^1 \!d \lamBar\, \Tr \left[ N G (\lamBar) \right] \, , 
\ee
where $G(\lamBar) = X^{-1}(\lamBar)$.
The computation of $G(\lamBar)$ can be first performed in the flat space:
\be \label{prop_def_flat}
X_0 (\lamBar)\, G_0 (\lamBar)= \mathbb{1} \, , \quad\quad
X_0 (\lamBar) = X (\lamBar) \rvert_{\na_\m \rightarrow p_\m} \, ,
\ee
where the operator $X_0 (\lamBar)$ is the flat space version of $X(\lamBar)$.
One can use an ansatz for $G_0$ with arbitrary coefficients and then work out the solution for them.
For the kinetic operator \eqref{kinop_big_expr} the solution of the equation \eqref{prop_def_flat} is given by \eqref{propagator_flat_expression}.
Then, we replace the momenta vectors back with covariant derivatives in $G_0$ with arbitrary ordering and get
\be \label{def_Mop}
X (\lamBar)\, G_0 (\lamBar) \rvert_{p_\m \rightarrow \na_\m} = \mathbb{1} + M(\lamBar, \bar{\na},\bar{R}) \, ,
\ee
where $M$ is at least linear in curvatures.
Now $G(\lamBar)$ can be expressed as a geometric series in $-M$:
\be \label{curved_propagator_expansion_as_geometric_series}
G = G_0 \frac{\mathbb{1}}{\mathbb{1}+M} = G_0 \left[ \mathbb{1} - M + M^2 - \dots \right] \, .
\ee
This expansion is infinite, but in order to see the divergent contributions to the effective action in four space-time dimensions, we need to keep only the terms up to quadratic order in curvatures.
Then, for \eqref{kinop_XY_split} we have
\bear
& \Tr\Log \left[ X + Y \right ] 
\\
= & \,\Tr\Log X + \Tr\Log \left[ \mathbb{1} + Y X^{-1} \right] 
\\
= & \,\Tr\Log X_{\rm min.} + \int_0^1 \!d \lamBar\, \Tr \left[ NG(\lamBar) \right] + \Tr \left[ YG - \polov YGYG \right] + \dots
\label{Tr_log_expr_2}
\eear
In the second row, we used the identity 
\begin{equation}
\Tr \Log\, AB = \Tr \Log\, A + \Tr \Log\, B,
\end{equation}
which is valid for any (non-commuting) positive definite operators,\footnote{Strictly speaking, this identity can be proven for finite-dimensional matrices. In the infinite-dimensional case, nontrivial corrections to this identity may be required, depending on which regularization scheme is used \cite{Kontsevich:1994nc, Evans:1998pd}, see also \cite{Cognola:2014pha, Elizalde:1997nd,
Barvinsky:2024irk, Shapiro:2025pmn} for related discussions.
The implications of this issue deserve further investigation.} 
while in the last row, we used the trace cyclic identity.
We can further utilize the definition of $G(\lamBar)$ and the fact that the trace of unity does not contribute to logarithmic divergence to get
\be
\Tr \Big[ G(\lamBar) X(\lamBar) \Big]_{\logdivsubscript} = \Tr \Big[ \!\left. G \right\rvert_{\lamBar = 0} X_{\rm min.} \Big]_{\logdivsubscript} = \left. \Tr\, \mathbb{1} \right\rvert_{\logdivsubscript} = 0 \, .
\ee
This allows us to rewrite the term under the integral as
\be
\Tr \Big[ N \, G(\lamBar) \Big] = \frac{1}{\lamBar} \Tr \Big[\! \left(X(\lamBar) - X_{\rm min.} \right) G(\lamBar) \Big] = \frac{1}{\lamBar} \Tr \Big[ \mathbb{1} - X_{\rm min.} G(\lamBar) \Big] 
= \frac{1}{\lamBar} \Tr \Big[ G(\lamBar) X_{\rm min.} \Big].
\ee
With our simplifications, up to the second order in curvature and torsion, we can finally represent the desired one-loop logarithmic divergence as
\be \label{log_div_master_equation}
\Tr\Log F =  \Tr\Log X_{\rm min.} + \int_0^1 \frac{d \lamBar}{\lamBar} \Tr \left[ G(\lamBar) X_{\rm min.} \right] 
+ \Tr \left[ Y G_0 \!\left(\mathbb{1}-M-\polov YG_0 \right) \right]_{\lamBar=1} + \dots \, .
\ee

The functional trace involving a minimal operator of the Laplace type relates to the trace of the heat kernel
\cite{Fock:1937dy, Schwinger:1951nm}:
\be
\Tr\Log \D = - \int_0^\infty \frac{ds}{s} \Tr\, e^{-s\D} \, .
\ee
We define $\D = - \Box + E$, where $E$ does not contain derivatives.
Using the cut-off regularization, we see that its logarithmically divergent contribution is proportional to the $a_{d/2}$ heat kernel coefficient \cite{DeWitt:1964mxt} (for review, see \cite{Avramidi:2000bm, Vassilevich:2003xt, Buchbinder:2021wzv}):
\be
\label{tr_log_min_div}
\left. \Tr \Log \D \right\rvert_\logdivsubscript = - \frac{1}{\left(4\pi\right)^{d/2}} \Log\! \left( \frac{\L_{\text{\tiny UV}}^2}{\m^2} \right) \!\int\! d^d x \,\sqrt{g} \,\tr\, a_{d/2}(x)\, ,
\ee
where $\m$ is the running scale and $d$ is the number of space-time dimensions.
For traces involving derivatives, we use the off-diagonal heat kernel technique developed in \cite{Barvinsky:1985an, Groh:2011dw}, (see also \cite{Benedetti:2010nr, Knorr:2021slg}), which leads to the following compact expression \cite{Melichev:2023lpn}:
\be \label{tr_ders_with_inverse_del_log_div}
\Tr \left[\nabla_{(\m_1} \dots \nabla_{\m_N)} \frac{1}{\D^k} \right]_\logdivsubscript = \frac{1}{\left(4\pi\right)^{d/2}} \frac{1}{\Gamma(k)} \Log\! \left( \frac{\L_{\text{\tiny UV}}^2}{\m^2} \right) \!\int\! d^d x\,\sqrt{g} \;\tr\, K_{(\m_1 \dots \m_N)}^{\left( \frac{d}{2} + \lfloor N/2 \rfloor - k \right)}(x) \, 
\ee
for positive integer $k$, where $\lfloor N/2 \rfloor$ is the floor function and $K$ are local curvature structures given by \eqref{K_heat_kernel_invariants}, while 
\be \label{tr_ders_no_inverse_del_log_div}
\Tr \left[\nabla_{(\m_1} \dots \nabla_{\m_N)} \right]_\logdivsubscript = 0 \, .
\ee
The logarithmic divergences, which are our main focus, are universal, meaning, independent of the regularization scheme \cite{Salam:1951sj}.
If one uses dimensional regularization instead, the obtained results would correspond to ours as \cite{Teixeira:2020kew, Buchbinder:2021wzv}:
\be \label{cut-off-dim-reg_correspondence}
\Log \frac{\L^2}{\m^2} \longleftrightarrow \frac{\left( \m^2 \right)^{\frac{d}{2}-2}}{2-\tfrac{d}{2}}\, .
\ee
The expression for the Faddeev--Popov ghost operator can be obtained in the standard way from \eqref{gauge_chi_def} assuming $c=0$:
\be
\D_{\text{gh}} = - \d^\m_\n \Box - (1+2b) \na^\m \na_\n - R^\m{}_\n \, .
\ee
It produces the logarithmically divergent contribution that reads
\be \label{log_div_ghost}
- \frac{11}{180} R_{\m\n\r\l}^{\, 2}
+ \frac{67+44b-8b^2}{360(1+b)^2} R_{\m\n}^{\, 2}
+ \frac{23+40b+20b^2}{144(1+b)^2} R^2 \, .
\ee
At the same time, the third ghost operator \eqref{3rd_gh_op_def} with our choice of gauge fixing parameters \eqref{Gauge_Choice_Percacci} gives
\be \label{log_div_3rd_ghost}
- \frac{11}{180} R_{\m\n\r\l}^{\, 2}
+ \frac{43}{90} R_{\m\n}^{\, 2}
- \frac{1}{9} R^2\, .
\ee

When computing the second term in \eqref{log_div_master_equation}, one has to perform a multiplication of pseudo-differential operators.
In particular, $G = X^{-1}$ will contain the inverse powers of the Laplace operator.
Such a product will contain contributions from the commutators, and here we explain a simple way to account for them.
For an arbitrary differential operator $O$ and function $f$, using the Laplace transform, one can derive the following formula \cite{Groh:2011dw}:
\be \label{commutator_with_box_from_Laplace}
\big[ f\ofbox\, , O \big] = \sum_{n=1}^\infty \frac{1}{n!} \, \big[ \Box\, , O \big]_n \, f^{(n)} \ofbox \, ,
\ee
where $[.\, , .]_n$ are the left-side nested commutators, $[\Box\, , O]_{n+1}=[\Box\, , [\Box\, ,O]_n]$, $[\Box\, ,O]_0 = O$.
In practice, it is convenient to hold an agreement that all pseudo-differential operators always contain (functions of) boxes only at the very right (or, alternatively, at the very left) of the expression.
That means that such operators are written in the form
\be
\sum_i {\d \varphi}{}_a O_i{}^a{}_b (\na) f_i \ofbox \d \varphi^b \, .
\ee
Omitting the summation, we can represent a product of two such operators $O_L (\na) f_L \ofbox$ and $O_R (\na) f_R \ofbox$ also in the same form using \eqref{commutator_with_box_from_Laplace} as
\bear
& 
O_L (\na) f_L \ofbox \times O_R (\na) f_R \ofbox = O_L (\na) O_R (\na) f_L \ofbox f_R \ofbox
\\&
+ \sum_{n=1}^{\infty} \frac{1}{n!} O_L (\na) [\Box \, , O_R (\na)]_n \;f_L^{(n)} \ofbox f_R \ofbox 
\, .
\eear
Each commutator produces curvature tensors, and so only the first two terms of this expansion may contribute to the one-loop divergences (in four dimensions).

For the quadratic gravity, the Euclidean action reads 
\be
\label{action_4DG_Weyl}
S_{\rm 4DG} = \frac{1}{2} \int d^4 x\;\sqrt{g} \left[ m_P^2 \left( 2 \Lambda - R \right)
+ \frac{1}{\l} C^{2} 
- \frac{1}{3\x} R^2 
+ \frac{1}{\r} E_{\rm GB} 
\right] \, .
\ee
The procedure described above yields the following beta functions \footnote{Note that in this paper the definitions of marginal couplings are slightly changed with respect to the nomenclature of \cite{deBerredo-Peixoto:2004cjk, Codello:2008vh, Percacci:2017fkn}, namely, we rescale $\x \rightarrow - \x/2$, and $\r \rightarrow - \r/2$, while $\l$ stays unchanged.
}
\bear \label{betas_4DG_lxr}
(4\pi)^2 \b_{\l}^{\rm g} &= - \frac{133}{10} \l^2 \, , 
\\
(4\pi)^2 \b_{\x}^{\rm g} &= \frac{5}{6} \left(2 \l^2 + 6 \l \x + \x^2 \right) \, ,
\\
(4\pi)^2 \b_{\r}^{\rm g} &= - \frac{392}{45} \r^2 \, ,
\eear
independently of the gauge choice \cite{Fradkin:1981iu,
Avramidi:1985ki, 
deBerredo-Peixoto:2004cjk}.
The choice of signs in \eqref{action_4DG_Weyl} is dictated by the condition of the absence of tachyons, which demands $\l>0$ and $\x>0$.
One can see that while the coupling $\l$ is asymptotically free, the coupling $\x$ is not.
In the next section, we will explore how torsion loops contribute to the running of these couplings.

\section{\label{sec:betas}Beta Functions}

We will proceed by considering subclasses of \eqref{action_decomposed} obtained by imposing kinematic constraints on torsion.
In order not to overwhelm the main text, we present only those results that are most relevant to our discussion.
Some of the remaining beta functions can be found in the App. \ref{sec:app.results}
and in the attached files in \emph{Mathematica} and text formats.
We work in the Euclidean regime, where the actions below differ from \eqref{action} and \eqref{action_decomposed} by an overall sign.
We will focus on the running of the marginal couplings in front of the curvature squared terms, which modify the expressions \eqref{betas_4DG_lxr}.
Similar to the metric 4DG case, massive couplings $m_p$, $\L$, and torsion masses do not contribute and can be dropped out of the calculation.

\subsection{\label{run_tor_vect}Vectorial and Axial Torsion}

Assuming that only the vector component of torsion can propagate while others are suppressed or absent leads to the following Euclidean action:
\bear \label{action_vectorial}
S_E^{\text{vect}} &= \polov \int d^4 x \sqrt{g} \Big[
m_p^2 \left( 2 \L - R \right)
+ \frac{1}{\l} C^2
- \frac{1}{3\x} R^2 + \frac{1}{\r} E_{\rm GB} 
\\&
- d_1\, T_{\m} \Box T^{\m} 
- d_2\, T_{\m} \na_{\m} \na^{\n} T^{\n} 
+ r_1\, R\,\na_{\m} T^{\m} 
+ m_{\tv}^2\, \tv_{\m} \tv^{\m} 
\Big] \, .
\eear
This theory has been studied in
\cite{Piva:2023noi} in the case $r_1=0$.
Apart from the proper vector, it also propagates a particle of spin zero that is either a ghost or purely virtual, depending on the quantization prescription \cite{Anselmi:2020opi}.
Note that we would have obtained the same action if instead of vectorial torsion we considered vectorial nonmetricity (also known as the Weyl vector $Q_\r \equiv g_{\m\n} D_\r g^{\m\n}$, where $D$ stands for the derivative with full connection).
Weyl geometry and its phenomenological consequences have been extensively studied in \cite{Ghilencea:2020rxc,
Ghilencea:2020piz,
Condeescu:2023izl,
Condeescu:2024nbs,
Ghilencea:2024usf}.
For the motivations to introduce the mixing term between the curvature and torsion $R\na\tv$ in the context of Weyl gauge theory, see \cite{Barker:2024goa}.
For perturbative quantum effects, the exact geometric nature of $\tv_\m$ is irrelevant.
This rank-1 field is not tachyonic if
\be \label{no-tachyons-conditions-vect}
\frac{d_2}{d_1}>-1 ~~ \text{and} ~~\, \frac{d_4}{d_3} >-1 \, .
\ee

The beta functions of the gravitational couplings consist of the graviton contributions \eqref{betas_4DG_lxr} plus the contribution from the vectorial loop:
\be
\b_{\l} = \b_{\l}^{\rm g} + \b_{\l}^{\rm v} \, , \quad
\b_{\x} = \b_{\x}^{\rm g} + \b_{\x}^{\rm v} \, , \quad
\b_{\r} = \b_{\r}^{\rm g} + \b_{\r}^{\rm v} \, , \quad
\ee
where the contributions from the vectorial part to the gravitational counterterms are 
\begingroup
\allowdisplaybreaks
\begin{subequations}
\begin{align} 
\label{beta_vectorial_lam}
(4\pi)^2 \b_{\l}^{\rm v} &= \frac{37}{120} \l^2
-\frac{2 d_1^2 \l^2}{3 \left(4 d_1+4 d_2 + 3 \x r_1^2 \right)^2} \, ,
\\
\label{beta_vectorial_xi}
(4\pi)^2 \b_{\x}^{\rm v} &= \x^2 \frac{16 \left(4 d_1^2 - 8 d_1 d_2 - 7 d_2^2 \right) + 72 \left(2 d_1 - d_2\right) r_1^2 \x + 81 r_1^4 \x^2}{24 \left(4 d_1 + 4 d_2 + 3 r_1^2 \x \right)^2} \, ,
\\
\label{beta_vectorial_rho}
(4\pi)^2 \b_{\r}^{\rm v} &= \frac{67 \r^2}{360}-\frac{ d_1^2 \r^2}{3 \left(4 d_1+4 d_2 + 3 \x r_1^2 \right)^2} \, .
\end{align}
\end{subequations}
\endgroup
For the $\mathbb{Z}_2$ symmetric actions with $r_1=0$ they have been computed in \cite{Piva:2023noi}.\footnote{In order to make contact with the results of \cite{Piva:2023noi} one needs to make the following replacements: $d_2 \rightarrow \l_2-1$, $\eta_2 = -1$.}
One can see that in the limit $d_2=r_1=0$, the vector field produces contributions which are independent of $d_1$:
\be
\left.\b_{\l}^{\rm v}\right\rvert_{d_6=r_1=0} = \frac{1}{(4\pi)^2} \frac{4\l^2}{15} \, , \quad 
\left.\b_{\x}^{\rm v}\right\rvert_{d_6=r_1=0} = \frac{1}{(4\pi)^2}\frac{\x^2}{6} \, , \quad
\left.\b_{\r}^{\rm v}\right\rvert_{d_6=r_1=0} = \frac{1}{(4\pi)^2} \frac{13\r^2}{90} \, .
\ee

Vectorial and axial components of torsion have very different geometrical meanings.
Moreover, axial torsion can couple to fermions and lead to intriguing phenomenology \cite{Karananas:2024xja, Fabbri:2025xbm}.
However, as one can see in \eqref{action_decomposed}, the action in the case of axial torsion is exactly the same as for vectorial torsion, apart from the fact that one of the nonminimal terms is absent.
Therefore, the running due to the axial component will be identical to the vectorial contributions, with the replacement $d_1 \rightarrow d_3$, $d_2 \rightarrow d_4$, $r_1 \rightarrow 0$.

\subsection{\label{run_tor_pure}Hook-Antisymmetric Traceless Torsion}

The remaining component of torsion produces more fruitful kinetic mixing with two nonminimal terms, with Euclidean action
\bear \label{action_pure_tor}
S_E^{\text{tens}} &= \polov \int d^4 x \sqrt{g} \Big[
m_p^2 \left( 2 \L - R \right)
+ \frac{1}{\l} C^2
- \frac{1}{3\x} R^2 
+ \frac{1}{\r} E_{\rm GB} 
\\&
- d_5\, \tp_{\m\n\r} \Box \tp^{\m\n\r}
- d_6\, \tp_{\m\r\l} \na^{\m} \na_{\n} \tp^{\n\r\l} 
- d_7\, \tp_{\m\r\l} \na^{\m} \na_{\n} \tp^{\n\l\r}
\\&
+ r_2\, C_{\m\n\r\s} \na^{\m} \tp^{\n\r\s} 
+ m_{\tp}\, \tp_{\m\n\r} \tp^{\m\n\r}
\Big] \, .
\eear
The field $\tp$ contains two spin-2 particles (one physical, one ghost or purely virtual) as well as two spin-1 particles (one physical, one ghost or purely virtual).
The no-tachyon conditions are \cite{Anselmi:2020opi}:
\be \label{no-tachyons-conditions-pure}
-3 < d_6 \, ,\quad
-2-d_6<d_7<6+d_6 \, , \quad
\l>0 \, , \quad 
\x>0 \, .
\ee
Below, we normalized $d_5 = 1$.
This comes without a restriction of generality because our perturbative inversion of the kinetic operator \eqref{curved_propagator_expansion_as_geometric_series} demands $d_5 \neq 0$.
The beta functions of the gravitational couplings consist of the graviton contributions \eqref{betas_4DG_lxr} plus the contribution from the tensorial torsion loop:
\be \label{betas_tp}
\b_{\l} = \b_{\l}^{\rm g} + \b_{\l}^{\tp} \, , \quad
\b_{\x} = \b_{\x}^{\rm g} + \b_{\x}^{\tp} \, , \quad
\b_{\r} = \b_{\r}^{\rm g} + \b_{\r}^{\tp} \, , \quad
\ee
\begin{subequations}
\begin{align} 
\label{beta_lambda_pure_d6}
(4\pi)^2 &\left. \b_{\l}^{\tp} \right\rvert_{d_7=r_2=0} = 
\frac{\l^2}{4320 (2 + d_6)^2 (3 + d_6)^2 (6 + d_6)^2}
\Big( 20901888 + 41803776 d_6 \\& + 32664384 d_6^2 + 13040352 d_6^3 + 2910564 d_6^4 + 379716 d_6^5 + 29623 d_6^6 + 1215 d_6^7 \Big) \, , \nn
\\ \label{beta_xi_pure_d6}
(4\pi)^2 &\left. \b_{\x}^{\tp} \right\rvert_{d_7=r_2=0} = 
- \frac{\x^2}{864 (2 + d_6)^2 (3 + d_6)^2 (6 + d_6)^2}
\Big( 746496 + 3359232 d_6 \\&+ 3504384 d_6^2 + 1630368 d_6^3 + 389196 d_6^4 + 48300 d_6^5 + 2695 d_6^6 + 45 d_6^7
\Big) \, . \nn
\end{align}
\end{subequations}
For the minimal case $d_6=d_7=r_2=0$, the contributions of the hook-antisymmetric traceless torsion to the beta functions reduce to
\be
\b_{\l}^{\tp} = \frac{1}{(4 \pi)^2} \frac{56}{12} \l^2 \, , \quad
\b_{\x}^{\tp} = - \frac{1}{(4 \pi)^2} \frac{2}{3} \x^2 \, , \quad
\b_{\r}^{\tp} = \frac{1}{(4 \pi)^2} \frac{86}{45} \r^2 \, , \quad
\ee
in agreement with \cite{Piva:2021nyj}.
The more general results for the beta functions are presented in the App. \ref{sec:app.results} and in the attached files.

It would be tempting to also make a comparison between our results and the results presented in \cite{Martini:2023apm}.
This, however, does not seem possible at this stage.
The reason is that the authors of \cite{Martini:2023apm} have chosen to get rid of the vector mode within the hook-antisymmetric traceless component, which is achieved by a particular choice of couplings, and leads to the action being invariant under a gauge-like transformation.
This choice would render the principal part of our kinetic operator degenerate, in the sense discussed in section \ref{subsec:Lags}.
Another discrepancy is due to the zero modes, which appear on compact manifolds such as a sphere, considered in \cite{Martini:2023apm}.

To perform the calculation, the \emph{xAct} collection of computer algebra packages, specifically \emph{xTensor} \cite{xAct:xTensor}, \emph{Invar} \cite{Martin-Garcia:2007bqa, Martin-Garcia:2008yei}, \emph{SymManipulator} \cite{xAct:SymManipulator}, and \emph{xTras} \cite{Nutma:2013zea} were used.
It was necessary to introduce small modifications of certain functions of \emph{xTras} to make them efficient for such involved computations and perform a parallelization.
In short, one has to ensure that time-consuming operations, such as \emph{Expand} and \emph{Simplify}, are applied only when strictly necessary.
Certain ideas have been learned from \cite{xAct:xParallel}, but the code presented there was not used.
It would be interesting to investigate whether using other computer algebra systems, such as \emph{Cadabra} \cite{Peeters:2007wn, 
Price:2022wlt,
Castillo-Felisola:2022uvp,
Castillo-Felisola:2022pud,
Price:2023pfc} can produce significant gains in computational efficiency.

\section{\label{sec:discussion}Discussion of the RG Flow}

In this paper, we have performed a computation of the logarithmic divergent contributions to the running of the four-derivative gravitational couplings $\l$, $\x$, and $\r$ for theories with propagating vectorial, axial, and hook-antisymmetric traceless torsion fields.
The theory under consideration can be seen as a generalization of the quadratic four-derivative gravity and is power-counting renormalizable.
Within the background field method, we considered backgrounds with arbitrary Riemann curvature and vanishing torsion.
This is sufficient to see the contributions of the metric and torsion fluctuations to the running of the curvature squared terms.
The logarithmic divergences are isolated using the cut-off scheme, which is equivalent to the $1/\epsilon$ pole of dimensional regularization.
The obtained formulae are in agreement with several results previously known in the literature.

Let us first comment on the physical applicability.
Depending on the choice of couplings, the torsional degrees of freedom can contain tachyons.
We assume that the conditions for the absence of tachyons \eqref{no-tachyons-conditions-vect} and \eqref{no-tachyons-conditions-pure} are satisfied.
As discussed in sections \ref{sec:intro} and \ref{subsec:Lags}, even in the absence of tachyons, these fields are still unstable if quantized with Feynman prescription at infinitely high energies.
Instead, one needs to either consider energies below the torsion masses $m_\tv$, $m_\ta$, $m_\tp$ or quantize them as purely virtual particles following the procedure developed in \cite{Anselmi:2018kgz, 
Anselmi:2018ibi, 
Anselmi:2020opi, 
Piva:2021nyj, 
Piva:2023noi}.
If the reader finds both of these possibilities unconvincing, then the results of this paper shall be treated as a first step to the computations of a theory where only fully physical torsional degrees of freedom propagate.
For examples of such theories, we refer to \cite{Mikura:2023ruz} and postpone further studies of these realizations to subsequent works.

\begin{figure}[!htbp]
\centering
  \begin{minipage}{0.43\textwidth}
  \centering
    \includegraphics[width=\textwidth]{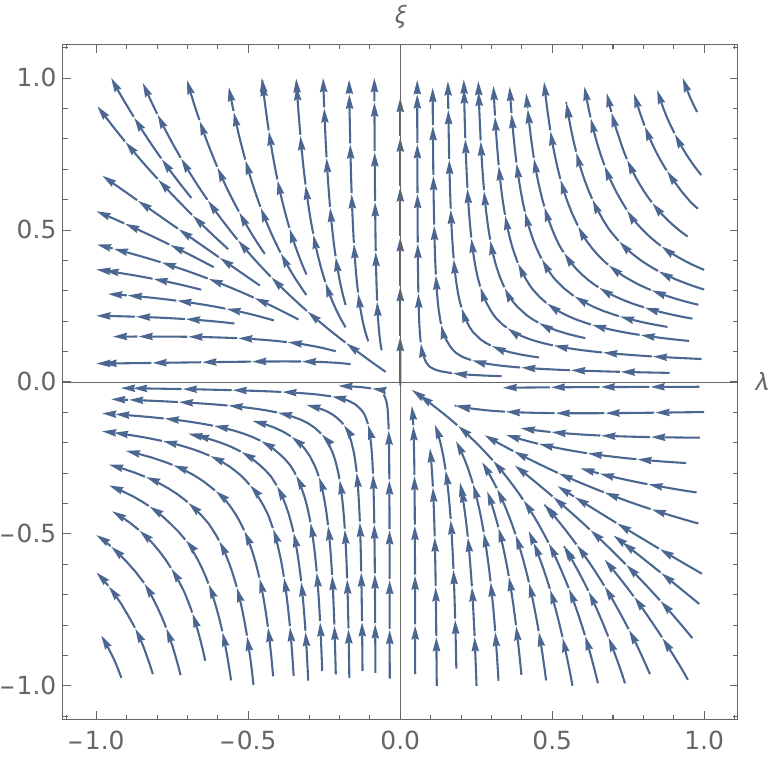}
    \caption{RG flow in pure metric case of the four-derivative quadratic gravity.}
  \label{fig:flow_4DG}
  \end{minipage}
  \hfill
  \begin{minipage}{0.43\textwidth}
  \centering
    \includegraphics[width=\textwidth]{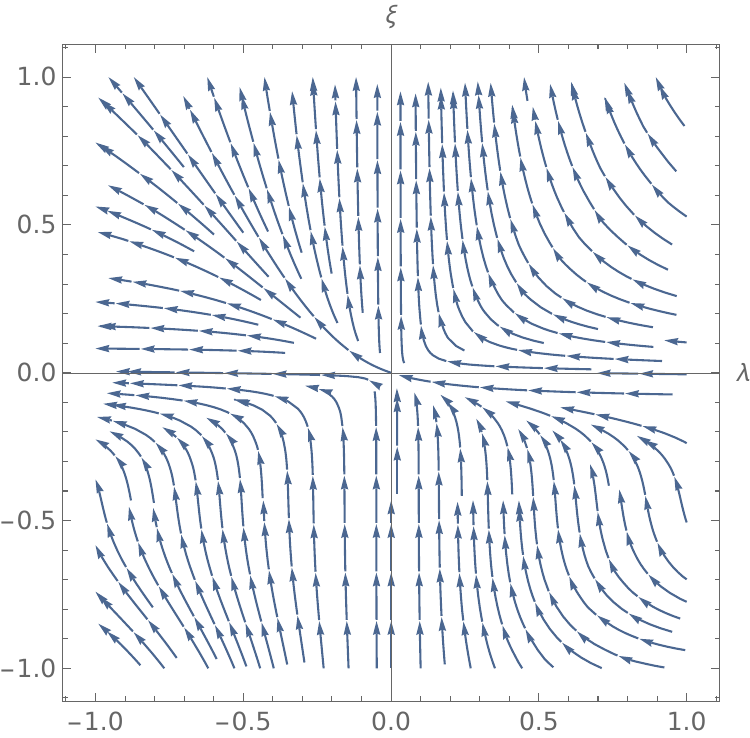}
    \caption{RG flow in 4DG with torsion vector, $d_1=1,~ d_2=r_1=-0.8$.}
  \label{fig:flow_mtg_d1eq1d2eqm08r1eqm08}
  \end{minipage}
\end{figure}

Let us discuss what can be learned from the obtained formulae.
Firstly, let us consider the four derivative gravity \eqref{action_4DG_Weyl}, which leads to the beta functions \eqref{betas_4DG_lxr}.
The Gau\ss--Bonnet contribution, that comes with the coupling $\r$, does not contribute to the spectrum and can be omitted.
The other two couplings $\l$ and $\x$ must be positive to avoid tachyons.
We display their RG flow in figure \ref{fig:flow_4DG}.
One can see that the flow lines inside the first quarter move away from the free fixed point.
This represents a well-known tension between the asymptotic freedom and the absence of tachyons \cite{Salvio:2014soa, Salvio:2018crh} (see, however, \cite{Buccio:2024hys} for the discussion of a different definition of the asymptotic freedom and a related computation).
The picture is also very similar when vectorial (or axial) torsion propagates on top of usual degrees of freedom, see figure \ref{fig:flow_mtg_d1eq1d2eqm08r1eqm08}.
This fact was already anticipated in \cite{Piva:2023noi}, where $\mathbb{Z}_2$-symmetric Lagrangian ($r_1=0$) was considered.
Here we confirm those findings and extend them for $r_1 \neq 0$.

\begin{figure}[!htbp]
  \centering
  \begin{subfigure}{0.43\textwidth}
    \centering
    \includegraphics[width=\textwidth]{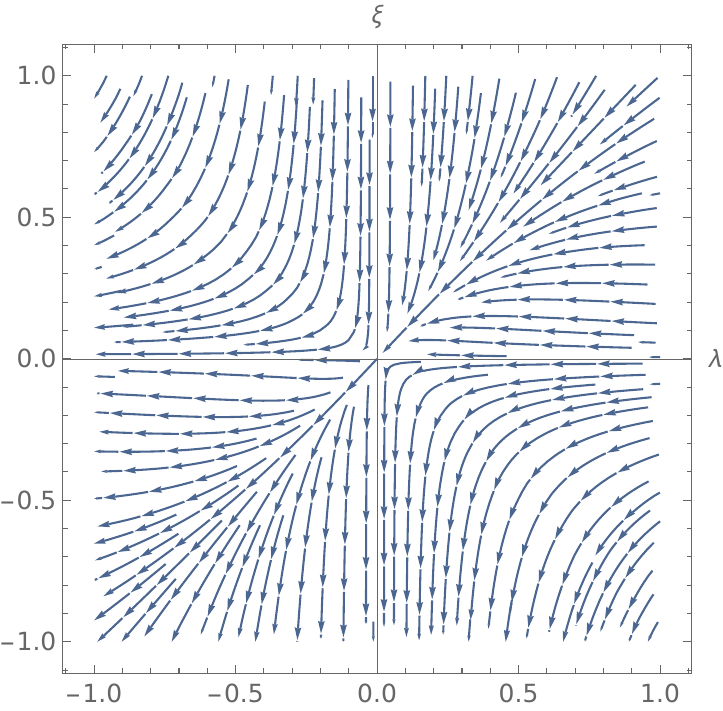}
    \caption{$d_6=1/2,~ d_7=r_2=0.$}
  \label{fig:flow_mtg_b6eq05b7eq00r2eq00}
  \end{subfigure}
  \hfill
  \begin{subfigure}{0.43\textwidth}
    \centering
    \includegraphics[width=\textwidth]{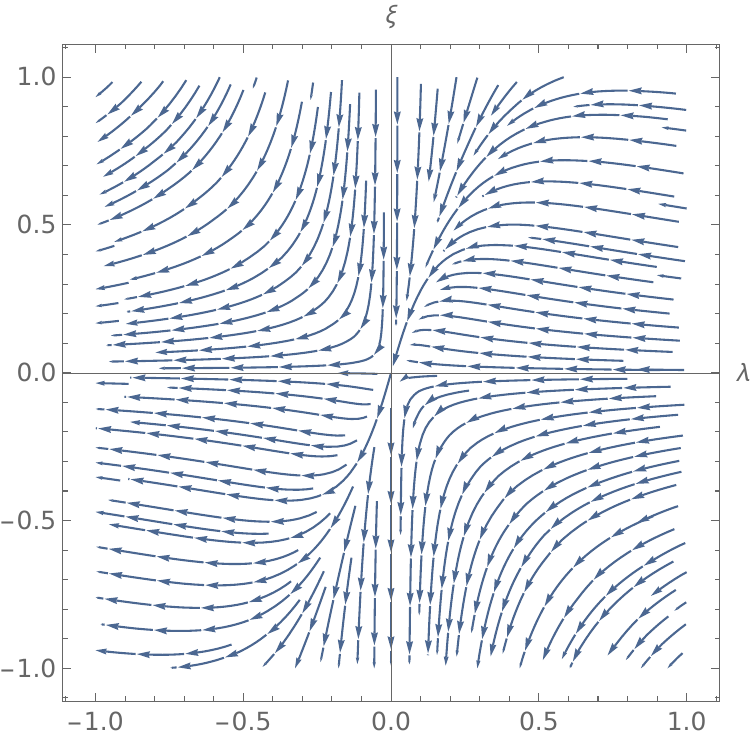}
    \caption{$d_6=d_7=1/3,~ r_2=1/2.$}
  \label{fig:flow_mtg_b6eq13b7eq13r2eq12}
  \end{subfigure}
  \caption{RG flow in gravity with propagating hook-antisymmetric traceless torsion depending on the choice of the remaining couplings. The tachyon-free sector $\l>0,\; \x>0$ flows towards the free fixed point $\l=\x=0$.}
  \label{fig:RG-flow_mtg}
\end{figure}

The situation is different when we consider the hook-antisymmetric traceless component of torsion.
The beta function of the coupling $\x$, shown in \eqref{betas_tp}, \eqref{beta_xi_pure_d6}, can change sign for sufficiently high values of the coupling $d_6$ (although perfectly within the tachyon-free region).
This leads to the change in the direction of the RG flow, displayed in the figure \ref{fig:RG-flow_mtg}.
One can see that the lines inside the first quarter flow towards the free fixed point, which corresponds to both $\b_\l$ and $\b_\x$ being negative for sufficiently small values of the couplings, rendering them asymptotically free.

\begin{figure}[!htbp]
\centering
  \begin{minipage}{0.48\textwidth}
  \centering
    \includegraphics[width=\textwidth]{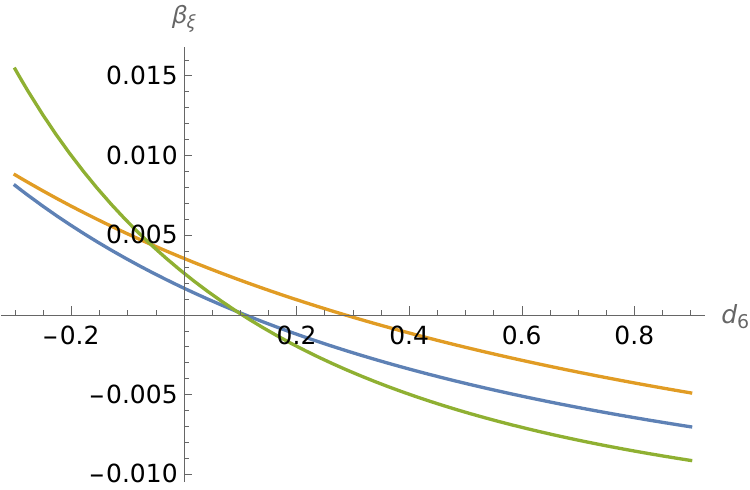}
    \caption{The values of the beta function of $\x$ coupling at $\x=0.1,~ \l=r_2=0$ as a function of $d_6$.
    The three lines correspond to $d_7=0$ (blue), $d_7=1/2$ (orange),  $d_7=-1/2$ (green).  
    The change of sign observed when passing to higher values of $d_6$ is related to the qualitative change of the flow from the form as the those represented in figures \ref{fig:flow_4DG} and \ref{fig:flow_mtg_d1eq1d2eqm08r1eqm08} to the form represented in figures \ref{fig:RG-flow_mtg}.
    }
    \label{fig:betas_tp_sign_change}
  \end{minipage}
  \hfill
  \begin{minipage}{0.39\textwidth}
  \centering
    \includegraphics[width=\textwidth]{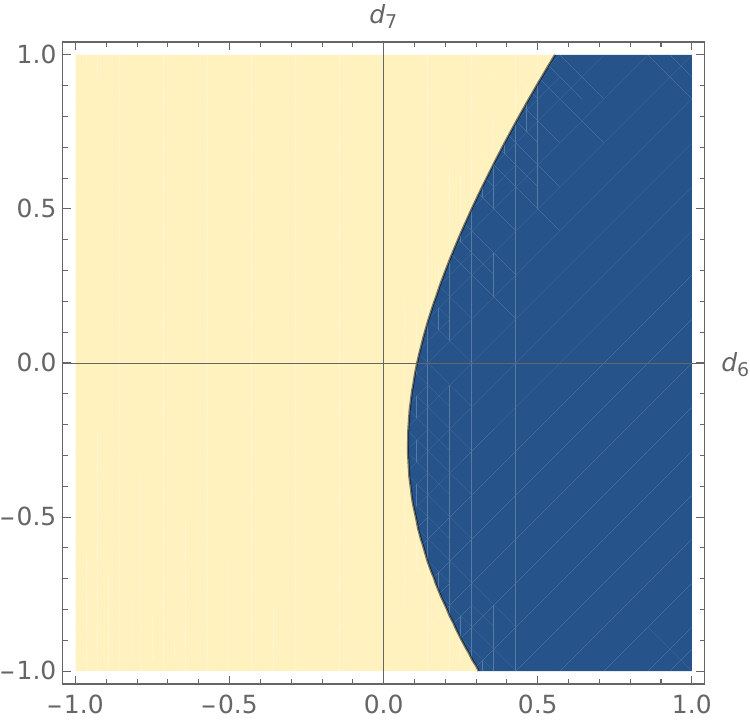}
    \caption{Sign of $\b_\x$ at the positive values of $\x$, with $\l=0$, depending on the values of the couplings $d_6$ and $d_7$. In the blue region, $\b_\x<0$ and observe asymptotic freedom. In the yellow region, $\b_\x>0$. The value of $r_2$ is irrelevant.}
  \label{fig:tp_asymptotic_freedom_landscape}
  \end{minipage}
\end{figure}

It was found that, depending on the values of $d_7$, and, especially, $d_6$, one observes that the full first quarter either flows away from the free fixed point or towards it.
It is of interest, therefore, to examine the exact moment when the RG flow switches its direction.
To this end, it is sufficient to consider the value of the beta function of the coupling $\x$ at the positive part of the axis $\l=0$.
The values of $d_7$ within the region $[-1;1]$ were considered, and it was found that it is always possible to find sufficiently high values of $b_6$ that lead to asymptotic freedom of the gravitational sector.
The results can be seen in figures \ref{fig:betas_tp_sign_change} and \ref{fig:tp_asymptotic_freedom_landscape}.

We point out that the calculation performed so far does not yet allow us to study the flow of the couplings of the kinetic terms for torsion.
To that end, one needs to perform a similar computation, but on a background with generic torsion.
We postpone this to further work, for the main findings of this paper could already be deduced from the computation that has been performed.
However, due to the large freedom in the choice of the couplings, there is a hope to eventually find stable setups with propagating, self-interacting torsion, resulting in a fully asymptotically free or asymptotically safe quantum field theory of MAG.

\section{\label{sec:Acknowledgments}Acknowledgments}

The author is indebted to Kevin Falls, Alexey Koshelev, Alexander Ochirov, Gregorio Paci, Dario Sauro, Anna Tokareva, and Yegor Zenkevich for multiple helpful conversations, as well as Roberto Percacci for useful suggestions and comments upon reading an earlier draft of the paper.

\begin{appendix}

\section{\label{sec:app.results}Results for the beta functions}

We present the general result for the running of the gravitational couplings $\l$ and $\x$ for the case with all nonminimal terms of the hook-antisymmetric traceless torsion.
The answer (with normalization of the minimal coupling $d_5=1$) is given by \eqref{betas_tp} with the gravitational contributions being \eqref{betas_4DG_lxr} as usual and
\begingroup
\allowdisplaybreaks
\begin{align*}
\label{beta_lambda_d67r2}  \numberthis
&(4\pi)^2 \b_\l^{\tp} = \frac{\l^2}{34560 (3 + 
 d_6)^2 (6 + d_6 - d_7)^2 (64 + 32 d_6 + 32 d_7 - \l r_2^2)^2} \Big[
 9953280 d_6^7 \\&+ 8192 d_6^6 (29623 + 360 d_7) - 
 24576 d_6^5 (-126572 - 7880 d_7 + 1095 d_7^2) - 
 8192 d_6^4 (-2910564 \\&- 348804 d_7 + 57401 d_7^2 + 720 d_7^3) + 
 24576 d_6^3 (4346784 + 816672 d_7 - 133208 d_7^2 - 12880 d_7^3 \\&+ 
 975 d_7^4) + 
 8192 d_6^2 (32664384 + 9183456 d_7 - 1247976 d_7^2 - 387144 d_7^3 + 
 34573 d_7^4 + 360 d_7^5) \\&- 
 24576 d_6 (-13934592 - 5666112 d_7 + 457632 d_7^2 + 431328 d_7^3 - 
 23916 d_7^4 - 5000 d_7^5 + 285 d_7^6) \\&- 
 73728 (-2322432 - 1375488 d_7 + 77184 d_7^2 + 87264 d_7^3 + 
 27612 d_7^4 - 10020 d_7^5 + 755 d_7^6) \\&+ 
 46080 (3 + d_6)^2 (6 + d_6 - d_7)^2 (8 + 6 d_6 + d_6^2 + 6 d_7 + 
 2 d_6 d_7 + d_7^2) r_2^2 \x 
\\&
+ \l \Big\{-1536 (555 d_6^6 - 3 d_6^5 (-4492 + 95 d_7) - 
 d_6^4 (-143565 - 5809 d_7 + 1110 d_7^2) \\&+ 
 2 d_6^3 (411219 + 59250 d_7 - 11573 d_7^2 + 285 d_7^3) + 
 d_6^2 (2618568 + 656514 d_7 - 127446 d_7^2 \\&- 3544 d_7^3 + 
 555 d_7^4) - 
 d_6 (-4318272 - 1465776 d_7 + 199494 d_7^2 + 69684 d_7^3 - 
 9670 d_7^4 + 285 d_7^5) \\&- 
 3 (-966816 - 331776 d_7 - 50040 d_7^2 + 74742 d_7^3 - 
 12915 d_7^4 + 755 d_7^5)) r_2^2 \\&- 
 2880 (3 + d_6)^2 (6 + d_6 - d_7)^2 (3 + d_6 + d_7) r_2^4 \x\Big\} 
\\&
 + \l^2 \Big\{72 (36288 + 235 d_6^5 + 378864 d_7 - 133632 d_7^2 + 
 15810 d_7^3 - 755 d_7^4 - 11 d_6^4 (-467 + 30 d_7) \\&- 
 2 d_6^3 (-20733 + 22 d_7 + 70 d_7^2) + 
 2 d_6^2 (72792 + 19791 d_7 - 4559 d_7^2 + 165 d_7^3) \\&- 
 d_6 (-202608 - 226800 d_7 + 65178 d_7^2 - 4780 d_7^3 + 
 95 d_7^4)) r_2^4 + 
 45 (3 + d_6)^2 (6 + d_6 - d_7)^2 r_2^6 \x \Big\}
\Big] \, ,
\end{align*}
\begin{align*}
\label{beta_xi_d67r2} \numberthis
& (4\pi)^2 \b_\x^{\tp} = 
\frac{5 \l^3 r_2^2 (128 + 64 d_6 + 64 d_7 - \l r_2^2)}{ 3 (64 + 32 d_6 + 32 d_7 - \l r_2^2)^2}
+ \frac{ 5 \l^2 r_2^2 (96 + 128 d_6 + 128 d_7 - 3 \l r_2^2) \x }{3 (64 + 32 d_6 + 32 d_7 - \l r_2^2)^2}
\\& 
- \frac{\x^2}{864 (3 + d_6)^2 (6 + d_6 - d_7)^2 (64 + 32 d_6 + 32 d_7 - \l r_2^2)^2}\Big[
 1024 \Big\{2695 d_6^6 + 45 d_6^7 \\& - 
 3 d_6^5 (-16100 - 428 d_7 + 45 d_7^2) - 
 d_6^4 (-389196 - 22932 d_7 + 6191 d_7^2) + 
 3 d_6^3 (543456 \\&+ 43968 d_7 - 20360 d_7^2 - 856 d_7^3 + 
 45 d_7^4) + 
 d_6^2 (3504384 + 373248 d_7 - 257832 d_7^2 - 37944 d_7^3 \\&+ 
 4297 d_7^4) - 
 3 d_6 (-1119744 - 124416 d_7 + 146592 d_7^2 + 47424 d_7^3 - 
 4260 d_7^4 - 428 d_7^5 \\&+ 15 d_7^6) - 
 9 (-82944 - 27648 d_7 + 78912 d_7^2 - 14400 d_7^3 + 8388 d_7^4 - 
 1668 d_7^5 + 89 d_7^6) \Big\}  
\\& 
- 192 \l r_2^2 \Big\{30 d_6^6 - 15 d_6^5 (-113 + d_7) - 
 d_6^4 (-23625 + 167 d_7 + 60 d_7^2) + 
 2 d_6^3 (66591 + 2394 d_7 \\&- 1295 d_7^2 + 15 d_7^3) + 
 2 d_6^2 (167076 + 22635 d_7 - 10509 d_7^2 + 217 d_7^3 + 
 15 d_7^4) - 
 d_6 (-309744 \\&- 104544 d_7 + 39798 d_7^2 + 4668 d_7^3 - 
 895 d_7^4 + 15 d_7^5) - 
 3 (-15552 + 19872 d_7 - 25560 d_7^2 \\&+ 13290 d_7^3 - 1971 d_7^4 + 
 89 d_7^5)\Big\} + 
 9 \l^2 r_2^4 \Big\{-60264 + 15 d_6^5 + 61776 d_7 - 19728 d_7^2 + 2274 d_7^3 \\& - 89 d_7^4 - 5 d_6^4 (-147 + 4 d_7) - 
 2 d_6^3 (-4155 + 297 d_7 + 5 d_7^2) + 
 2 d_6^2 (15912 - 933 d_7 \\&- 253 d_7^2 + 10 d_7^3) - 
 d_6 (-25056 - 13968 d_7 + 5838 d_7^2 - 454 d_7^3 + 
 5 d_7^4) \Big\} 
 \Big] \, .
\end{align*}
\endgroup
The expression for the beta function of the coupling $\r$ is of no big use, so we present it here only with the choice $d_5=1$ and $d_7=r_2=0$:
\bear
(4\pi)^2 \left. \b_{\r}\right\rvert_{d_7=r_2=0} &= 
\frac{\r^2}{2160 (2 + d_6)^2 (3 + d_6) (6 + d_6)^2}
\Big( 1783296 + 2972160 d_6 \\&+ 1947168 d_6^2 
+ 637008 d_6^3 + 105132 d_6^4 + 7928 d_6^5 + 225 d_6^6 \Big) \, .
\eear
All expressions for the general case are available in the attached files.

\section{\label{sec:app.relations}Some intermediate relations}

\subsection{\label{sec:app:Lags}Lagrangians and Decomposition of Torsion}

When we directly substitute \eqref{torsion_decomposition_def} into the original action \eqref{action}, the resulting expression will contain many more terms than the original one.
They, however, can all be brought into the neat form \eqref{action_decomposed} when one makes proper use of the properties \eqref{tensorial_torsion_properties}.
Below, we display several relations that are straightforward to prove.
For the mass terms, we have
\bear
& T^{\a\b\g} T_{\a\b\g} = \frac{2}{3} \tv_\m \tv^\m - \frac{1}{6} \ta_\m \ta^\m + \tp_{\a\b\g} \tp^{\a\b\g} \\
& T^{\a\b\g} T_{\a\g\b} = \frac{1}{3} \tv_\m \tv^\m + \frac{1}{6} \ta_\m \ta^\m + \frac{1}{2} \tp_{\a\b\g} \tp^{\a\b\g} \, .
\eear
For the kinetic terms, we need to generalize these expressions for non-contracted indices, such as
\begingroup
\allowdisplaybreaks
\begin{align*}
\label{decomposed_torsion_useful_relations}
& \tp_{\l\m\n} \tp^{\l\n\m} = \frac{1}{2} \tp_{\a\b\g} \tp^{\a\b\g} \, , 
\\ \numberthis
& \tp^{\l\m\n} \tp_{\m\r\n} = \frac{1}{2} \tp^{\m\l\n} \tp_{\m\r\n} \, , 
\\
& \tp^{\l\m\n} \tp_{\r\m\n} - \tp^{\l\m\n} \tp_{\r\n\m} - \tp_{\m}{}^{\l}{}_{\n} \tp_{\m\r\n} = 0 \, . 
\end{align*}
\endgroup
By introducing the Levi-Civita tensor
\be
\label{Levi_Civita_tensor_def}
\lctens^{\m\n\r\l} = \sqrt{g}\; \epsilon^{\m\n\r\l} \, ,
\ee
we can obtain many more tensorial identities that involve derivatives.
We will list some of them here:
\bear
& \lctens_{\a\b\d\z} \ta^\a (\na^\z \na_\g \tp^{\b\g\d}) = 2 \lctens_{\a\g\d\z} \ta^\a ( \na^\z \na_\b \tp^{\b\g\d} ) + \dots \, ,
\\ & 
\tp_{\l\m\n} \Box \tp^{\l\n\m} = \frac{1}{2} \tp_{\l\m\n} \Box \tp^{\l\m\n} + \dots \, .
\eear
where the ellipses denote terms of higher than quadratic order in torsion.
The second Bianchi identity yields
\be
R^{\m\n} \na_\m \tv_\n = \frac{1}{2} R \na_\m \tv^\m \, .
\ee
The map between the coupling of \eqref{action} and \eqref{action_decomposed} is 
\begingroup
\allowdisplaybreaks
\begin{align*} \label{decomposed-couplings-map}
&
\a = \frac{1}{6\l} - \frac{1}{6\x} - \frac{1}{2\r} \, , \quad
\b = - \frac{1}{\l} + \frac{2}{\r} \, , \quad
\g = \frac{1}{2\l} - \frac{1}{2\r} \, , \quad
\\
&
m_1 = -2 m_\ta^2 + \frac{2}{3} m_\tp^2 \, , \quad
m_2 = 4 m_\ta^2 + \frac{2}{3} m_\tp^2 \, , \quad
m_3 = m_\tv^2 - \frac{2}{3} m_\tp^2 \, , \quad
\\ &  \numberthis
\t = r_1 + \frac{r_2}{12} \, , \quad
\s = -\frac{r_2}{2} \, , \quad
b_1 = - 2 d_3 - 2 d_4 - \frac{2}{3} d_5 \, , 
\quad
b_2 = 4 d_3 + 4 d_4 - \frac{2}{3} d_5 \, , \quad
\\&
b_3 = d_1 + \frac{2}{3} d_5 - \frac{1}{9} d_6 - \frac{1}{3} d_8 \, , \quad 
b_4 = 4 d_4 - \frac{5}{9} d_6 - \frac{4}{9} d_7 - \frac{2}{3} d_9 \, , \quad
\\&
b_5 = - 4 d_4 - \frac{4}{9} d_6 - \frac{5}{9} d_7 + \frac{2}{3} d_9 \, , \quad
b_6 = 2 d_4 - \frac{1}{9} d_6 + \frac{1}{9} d_7 + \frac{2}{3} d_9 \, , \quad
\\&
b_7 = - 8 d_4 - \frac{2}{9} d_6 + \frac{2}{9} d_7 - \frac{2}{3} d_9 \, , \quad
b_8 = \frac{2}{3} d_6 + d_8 \, , \quad
b_9 = d_2 + \frac{4}{9} d_6 + \frac{1}{3} d_7 + \frac{1}{3} d_8 \, . 
\end{align*}
\endgroup
The reverse relations are
\begingroup
\allowdisplaybreaks
\begin{align*}
\label{decomposed-couplings-map-reverse}
&\l = \frac{1}{\b+4\g} \, , \quad 
\x =\frac{-1}{2(3\a+\b+\g)} \, , \quad
\r = \frac{1}{\b+2\g} \, ,
\\&
m_\tv^2 = \frac{2}{3} m_1 + \frac{1}{3} m_2 + m_3 \, , \quad
m_\ta^2 = \frac{1}{6} \left( - m_1 + m_2 \right)  \, , \quad
m_\tp^2 = m_1 + \frac{1}{2} m_2 \, , \quad
\\& \numberthis
r_1 = \frac{\s}{6} + \t \, , \quad
r_2 = - 2\s  \, , \quad
d_1 = \frac{1}{9} (6  b_1+3  b_2+9  b_3+ b_4+2  b_6+ b_7+3  b_8) \, , 
\\& 
d_2 = \frac{1}{9} (2  b_4+3  b_5-2  b_6- b_7-3  b_8+9  b_9) \, , \quad 
d_3 = \frac{1}{18} (-3  b_1+3  b_2- b_4+ b_5- b_6+ b_7) \, , \quad
\\&
d_4 = \frac{1}{18} ( b_4- b_5+ b_6- b_7) \, , \quad 
d_5 = - \frac{1}{2} (2  b_1 + b_2) \, , \quad 
d_6 = - b_4-2  b_6- b_7 \, , \quad
\\&
d_7 = - b_5+2  b_6+ b_7 \, , \quad 
d_8 = \frac{1}{3} (2  b_4+4  b_6+2  b_7+3  b_8) \, , \quad 
d_9 = \frac{1}{6} (-2  b_4+2  b_5+4  b_6- b_7) \, .
\end{align*}
\endgroup

\subsection{\label{subsec:kinop_expression}Kinetic operator in flat space}

Using the gauge fixing of the form \eqref{S_gf} with generic coefficients, the second variation of the action, together with a gauge-fixing term, gives:
\begingroup
\allowdisplaybreaks
\begin{align*} \label{kinop_big_expr}
S_{,2}
&= \tfrac{\b+4\g}{2} {h}{}^{\m\n} \Box^2 h_{\m\n} 
+ \frac{ 4 \a + \b }{2} h{} \Box^2 h 
- \frac{ b^2 (1 + e - f) }{a} h{} \Box^2 h 
- \frac{ b^2 d m_0}{a} h{} \Box h 
\\&
+ \left( 2\a+\b+2\g + \frac{f-e}{a} \right) h{}^{\m\n} \na_\m \na_\n \na_\r \na_\l h^{\r\l}
- \frac{d m_0}{a} h{}^{\m\n} \na_\m \na_\r h_\n{}^\r
\\& \numberthis
- d_1 {\d \tv}^\m \Box \d \tv_\m 
- d_3 {\d \ta}^\m \Box \d \ta_\m 
- d_2 {\d \tv}^\m \na_\m \na_\n \d \tv^\n
- d_4 {\d \ta}^\m \na_\m \na_\n \d \ta^\n
\\&
- d_{5}{}{\d \tp}{}^{\m\n\r} \Box \d \tp_{\m\n\r} 
+ \frac{r_1}{2} \left( {\d \tv}{}^{\a } \na_\a \Box h - h \na_\a \Box \tv^\a \right)
\\&
- \left( \frac{4\a+\b}{2} + \frac{b (1+e-f)}{a} \right) h{}^{\m\n} \na_\m \na_\n \Box h
- \frac{bdm_0}{a} h{}^{\m\n} \na_\m \na_\n h
\\&
+ \frac{r_2}{8} \left( h{}^{\m\n} \na^\r \Box \d\tp_{\m\n\r} - {\d\tp}{}_{\m\n\r} \na^\r \Box h^{\m\n} \right)
- \left( \b+4\g + \frac{1}{a} \right) h{}^{\m\n} \na_\m \na_\r \Box h_\n{}^\r
\\&
- \left( \frac{4\a+\b}{2} + \frac{b(1+e-f)}{a} \right) h \na_\m \na_\n \Box h^{\m\n} 
- \frac{b d m_0}{a} h \na_\m \na_\n h^{\m\n} 
\\&
+ \frac{r_1}{2} \left( h{}^{\m\n} \na_\m \na_\n \na_\r \d\tv^\r - {\d\tv}{}^{\r} \na_\r \na_\n \na_\m h^{\m\n} \right)
\\&
+ \frac{r_2}{8} \left( {\d\tp}{}^{\a\b\g} \na_\m \na_\g \na_\b \na_\a h_\a{}^\m - h{}^{\a\m} \na_\a \na_\b \na_\g \na_\m \d\tp_\a{}^{\b\g} \right)
\\&
+ \frac{d_7-d_6}{3} {\d\tp}{}^{\a\m\n} \na_\r \na_\m \d\tp_{\a\n}{}^\r
- \left( d_6 + \frac{d_7}{2} \right) {\d\tp}{}^{\a\m\n} \na_\r \na_\n \d\tp_{\a\m}{}^\r
\\&
- \frac{d_8}{2} \d\tv^\a \na_\m \na_\n \d\tp_\a{}^{\n\m}
- \frac{d_8}{2} \d\tp^{\a\m\n} \na_\m \na_\n \d\tv_\a
\\&
- \frac{d_9}{2} \eta_{\a\m\n\r} \d\tv^\a \na^\r \na_\b \d\tp^{\b\m\n}
- \frac{d_9}{2} \eta_{\a\b\m\n} \d\tp_\r{}^{\m\n} \na^\r \na^\b \d\tv^\a
- \frac{d_9}{2} \eta_{\a\b\m\n} \d\tp^\m{}_\r{}^\n \na^\r \na^\b \d\tv^\a \, .
\end{align*}
\endgroup
Contributions proportional to the background curvature or torsion are omitted.
When multiplying operators that act on constrained fields (such as $\d \tp_\m{}^\n{}_\r$, that is traceless and hook-antisymmetric), it is crucial to enforce that the result does not bring us outside of that restricted field space.
This means that the metric \eqref{configuration_space_metric} is always understood as a projection operator, acting right after every multiplication.

\subsection{\label{sec:app:prop-flat}Inversion of kinetic operator in flat space}

We find the solution of the equation \eqref{prop_def_flat} for the most general Lagrangian \eqref{action_decomposed} as
\be \label{propagator_flat_expression}
\mathbb{G}_0 (\lamBar) _{\a\b\;\x\;\omega\;\g}{}^\d{}_{\eta\, ,}{}^{\m\n\;\s\;\t\;\r}{}_\l{}^\z = \frac{1}{\Box^4}
\begin{pmatrix}
G_0^{hh}{}_{\,\a\b\, ,}{}^{\m\n} \quad&\quad G_0^{h\tv}{}_{\,\a\b\, ,}{}^{\s} \quad&\quad 0 \quad&\quad G_0^{h\tp}{}_{\,\a\b\, ,}{}^\r{}_\l{}^\z 
\\
- G_0^{h\tv\, \m\n\, ,}{}_\x \quad&\quad G_0^{\tv\tv}{}_{\,\x\, ,}{}^{\s} \quad&\quad 0 \quad&\quad G_0^{\tv\tp}{}_{\,\x\, ,}{}^\r{}_\l{}^\z
\\
0 \quad&\quad 0 \quad&\quad G_0^{\ta\ta}{}_{\,\omega\, ,}{}^{\t} \quad&\quad G_0^{\ta\tp}{}_{\,\omega\, ,}{}^\r{}_\l{}^\z 
\\
G_0^{\tp h}{}_{\,\g}{}^\d{}_{\eta\, ,}{}^{\m\n} \quad&\quad G_0^{\tp \tv}{}_{\, \g}{}^\d{}_{\eta\, ,}{}^\s \quad&\quad G_0^{\tp\ta}{}_{\,\g}{}^\d{}_{\eta\, ,}{}^\t \quad&\quad G_0^{\tp\tp}{}_{\,\g}{}^\d{}_{\eta\, ,}{}^\r{}_\l{}^\z
\end{pmatrix} \, ,
\ee
where
\begingroup
\allowdisplaybreaks
\begin{align*}
\label{G0_subexpr} \numberthis
G_0^{hh}{}_{\,\a\b\, ,}{}^{\m\n} &= \left( p_1 \d^\m_{(\a} \d^\n_{\b)} + p_2 g_{\a\b} g^{\m\n} \right) \Box^2 
+ p_3 g^{\m\n} \na_\a \na_\b \na^2 
+ p_4 g_{\a\b} \na^\m \na^\n \na^2
\\& \nn
+ p_5 \d^{(\m}_{(\a} \na^{\n)} \na_{\b)}
+ p_6 \na_\a \na_\b \na^\m \na^\n \, , 
\\ 
G_0^{h\tv}{}_{\,\a\b\, ,}{}^{\s} &=
q_1 \d^\s_{(\a} \na_{\b)} \Box^2 
+ q_2 g_{\a\b} \na^\s \Box^2 
+ q_3 \na_\a \na_\b \na^\s \na^2 \, ,
\\ 
G_0^{h\tp}{}_{\,\a\b\, ,}{}^\r{}_\l{}^\z &= 
q_4 \d^{[\r}_{(\a} g_{\b)\l} \na^{\z]} \na^2 
+ q_5 g_{\a\b} \d^\z_\l \na^\r \na^2 
+ q_6 \d^{[\r}_{(\a} \na_{\b)} \d^{\z]}_\l \na^2 
+ q_7 \d^{\z}_\l \na_\a \na_\b \na^{\r]} 
+ q_8 \d^{[\r}_{(\a} \na_{\b)} \na_\l \na^{\z]} \, ,
\\ 
G_0^{\tv\tv}{}_{\,\x\, ,}{}^{\s} &=
k_1 \d^\s_\x \Box^2 
+ k_2 \na_\x \na^\x \na^2 \, , 
\\ 
G_0^{\tv\tp}{}_{\,\x\, ,}{}^\r{}_\l{}^\z &=
k_3 \d^{[\r}_\x \d^{\z]}_\l \Box^3
+ k_4 \d^{[\r}_\x \na^{\z]} \na_\l \Box^2 
+ k_5 \d^{[\z}_\l \na_\x \na^{\r]} \Box^2 \, ,
\\ 
G_0^{\ta\tp}{}_{\,\omega\, ,}{}^\r{}_\l{}^\z &=
k_{11} \eta^{[\r}{}_{\l\omega\a} \na^{\z]} \na^\a \Box^2 
+ k_{12} \eta^{\r\z}{}_{\omega\a} \na^\a \na_\l \Box^2 \, ,
\\ 
G_0^{\tp\ta}{}_{\,\g}{}^\d{}_{\eta\, ,}{}^\t & =
k_{13} \eta_{\g\eta}{}^\t{}_\a \na^\a \na^\d \Box^2 \, ,
\\ 
G_0^{\tp\tp}{}_{\,\g}{}^\d{}_{\eta\, ,}{}^\r{}_\l{}^\z & = 
k_{14} \d^\d_\g \d^\d_\eta \na^\z \na_\l \Box^2
+ k_{15} \d^\r_\g \g_{\eta\l} \na^\r \na^\z \Box^2
+ k_{16} \d^\r_\g \na^\d \na^\z \na_\eta \na_\l \Box 
+ k_{17} \d^\d_\g g^{\d\z} g_{\eta\l} \Box^3
\\
+ k_{18} g^{\d\r} g_{\g\l}& \na^z \na_\eta \Box^2
+ k_{19} \d^\d_\g \d^\r_\l \na^\z \na_\eta \Box^2
+ k_{20} \d^\r\g \d^\z_\eta \na^\d \na_\l \Box^2
+ k_{21} \d^\d_\g \d^\r_\eta \d^\z_\l \Box^3
+ k_{22} \d^\r_\eta \d^\z_\l \na_\g \na_\d \Box^2
\, .
\\
G_0^{\tv h}{}_{\,\a\b\,}{}^{\m\n} &= \left. G_0^{h \tv}{}_{\,\a\b\,}{}^{\m\n} \right\rvert_{pp} \, ,
\\
G_0^{\tp h}{}_{\,\g}{}^\d{}_{\eta\, ,}{}^{\m\n} &= \left. G_0^{h \tp}{}_{\,\m\n\, , \g}{}^\d{}_{\eta} \right\rvert_{\{ q_4, q_5, q_6, q_7, q_8 \} \rightarrow \{ q_9, q_{10}, q_{11}, q_{12}, q_{13} \} } \, ,
\\
G_0^{\ta \ta}{}_{\,\omega\, ,}{}^\t & = 
\left. G_0^{\tv\tv}{}_{\,\omega\, ,}{}^\t \right\rvert_{\{ k_1, k_2 \} \rightarrow \{ k_9, k_{10} \} } \, ,
\\ 
G_0^{\tp \tv}{}_{\, \g}{}^\d{}_{\eta\, ,}{}^\s &= \left. G_0^{\tv \tp\, \s}{}_{\,\g}{}^\d{}_{\eta} \right\rvert_{\{ k_3, k_4, k_5 \} \rightarrow \{ k_6, k_7, k_8 \} } \, .
\end{align*}
\endgroup
For the action with the vectorial component only \eqref{action_vectorial} we obtain the following solution for the inversion coefficients that enter \eqref{G0_subexpr}:
\bear
&
p_1 = 2 \l \, , \quad 
p_2 = -\frac{2 \left(2 d_1 (2 \l + \x ) + \lamBar \left(2 d_2 (2 \l + \x ) + 3 \l \lamBar r_1^2 \x \right)\right)}{3 l_1} \, , \quad 
p_3 = \frac{2 \l \lamBar^2 r_1^2 \x }{l_1} \, , \quad 
\\& 
p_4 = \frac{2 \l \lamBar^2 r_1^2 \x }{l_1} \, , \quad 
p_6 = \frac{4 \l^2 \lamBar^2 r_1^2}{l_1} \, , \quad 
q_2 = \frac{2 \lamBar r_1 \x }{l_1} \, , \quad 
q_3 = \frac{4 \l \lamBar r_1}{l_1} \, , \quad 
k_1 = -\frac{1}{d_1 } \, , \quad 
\\& 
k_2 = \frac{\lamBar \left(4 d_2 + 3 \lamBar r_1^2 \x \right)}{d_1 l_1} \, , \quad 
l_1 = 4 d_1 + \lamBar \left(4 d_2 + 3 \lamBar r_1^2 \x \right) \, , 
\eear
with all other coefficients equal to zero.
For the action with hook-antisymmetric traceless component only \eqref{action_pure_tor}, we have:
\begingroup
\allowdisplaybreaks
\begin{align*}
\label{solutioninverse_hooksector}
&
p_1 = \frac{64 \l (2 d_5+\lamBar (d_6+d_7))}{l_4 } \, , \quad 
p_2 = \frac{\lamBar \left(-32 d_6 (2 \l +\x )-32 d_7 (2 \l +\x )+\l \lamBar r_2 ^2 \x \right)-64 d_5 (2 \l +\x )}{192 d_5+3 l_2 \lamBar} \, , \quad 
\\&
p_3 = \frac{2 \l ^2 \lamBar^2 r_2 ^2}{192 d_5+96 d_6 \lamBar+96 d_7 \lamBar-3 \l \lamBar^2 r_2 ^2} \, , \quad 
p_4 = \frac{2 \l ^2 \lamBar^2 r_2 ^2}{192 d_5+96 d_6 \lamBar+96 d_7 \lamBar-3 \l \lamBar^2 r_2 ^2} \, , \quad 
\\&
p_5 = -\frac{4 \l ^2 \lamBar^2 r_2 ^2}{l_4 } \, , \quad 
p_6 = \frac{4 \l ^2 \lamBar^2 r_2 ^2}{192 d_5+96 d_6 \lamBar+96 d_7 \lamBar-3 \l \lamBar^2 r_2 ^2} \, , \quad 
q_4 = \frac{16 \l \lamBar r_2 }{l_4 } \, , \quad 
\\& \numberthis
q_5 = - q_7 = \frac{16 \l \lamBar r_2 }{192 d_5+96 d_6 \lamBar+96 d_7 \lamBar-3 \l \lamBar^2 r_2 ^2} \, , \quad 
q_8 = q_9 = - q_{13} = - \frac{16 \l \lamBar r_2 }{l_4 } \, , \quad 
\\&
k_{15} = \frac{2 \lamBar \left(d_5 \left(128 d_6+64 d_7-3 \l \lamBar r_2 ^2\right)+l_2 \lamBar (d_6-d_7)\right)}{d_5 l_3 l_4 } \, , \quad 
\\&
k_{16} = \frac{\lamBar^2 \left(18 d_5^2 \l r_2 ^2+12 d_5 d_7 l_2 +d_6 l_2 \lamBar (d_7-d_6)\right)}{d_5 l_3 l_4 (3 d_5+d_6 \lamBar)} \, , \quad 
k_{17} = -\frac{2}{d_5} \, , \quad 
\\&
k_{18} = \frac{2 l_2 \lamBar}{64 d_5^2+d_5 l_2 \lamBar} \, , \quad 
k_{20} = \frac{\lamBar (d_6-d_7)}{d_5 l_3 } \, , \quad 
k_{22} = \frac{d_6 \lamBar}{3 d_5^2+d_5 d_6 \lamBar} \, , \quad
\\&
q_1 = q_2 = q_3 = q_6 = q_{10} = q_{11} = q_{12} = k_1 = \dots = k_{14} = k_{19} = k_{21} = 0 \, , \quad 
\\&
l_2 = 32 d_6+32 d_7-\l \lamBar r_2 ^2 \, , \quad 
l_3 = 6 d_5+\lamBar (d_6-d_7) \, , \quad 
l_4 = 64 d_5+l_2 \lamBar \, .
\end{align*}
\endgroup

\subsection{\label{sec:app:heat_kernel_coeffs}Heat Kernel Expansion}

The first six invariants that enter \eqref{tr_ders_with_inverse_del_log_div} are \cite{Groh:2011dw}:
\begingroup
\allowdisplaybreaks
\begin{align*} \label{K_heat_kernel_invariants}
K^{(n)}(x) =&\, \overline{a_n} \, ,\\
K^{(n)}_{\m}(x) =&\, \overline{\na_\m a_n} \, ,\\
K^{(n)}_{(\m\n)}(x) =&\,
 -\frac{1}{2} g_{\m\n} \overline{a_n}
 + \overline{\na_{(\m}\na_{\n)} a_{n-1}} \, ,\\
K^{(n)}_{(\m\n\r)}(x) =&\, \numberthis
 -\frac{3}{2} g_{(\m\n} \overline{\na_{\r)} a_n}
 + \overline{\na_{(\m}\na_\n \na_{\r)} a_{n-1}} \, ,\\
K^{(n)}_{(\m\n\r\l)}(s) = &\,
 \frac{3}{4} g_{(\m\n}g_{\r\l)} \overline{a_n}
- 3 g_{(\m\n} \overline{\na_\r \na_{\l)} a_{n-1}}
+ \overline{\na_{(\m} \na_\n \na_\r \na_{\l)} a_{n-2}} \, ,\\
K^{(n)}_{(\m\n\r\l\a)}(x) = &\,
\frac{15}{4} g_{(\m\n}g_{\r\l} \overline{\na_{\a)}a_n} 
- 5 g_{(\m\n} \overline{\na_\r \na_\l \na_{\a)} a_{n-1}}
+ \overline{\na_{(\m}\na_\n \na_\r \na_\l \na_{\a)} a_{n-2}} \, , \\ 
K^{(n)}_{(\m\n\r\l\a\b)}(x) = &\, - \frac{15}{8} g_{(\m\n}g_{\r\l}g_{\a\b)} \overline{a_n} + \frac{45}{4} g_{(\m\n} g_{\r\l} \overline{\na_\a \na_{\b)} a_{n-1}}
- \frac{15}{2} g_{(\m\n} \overline{\na_\r \na_\lambda \na_\a \na_{\b)} a_{n-2}} \\
&
+ \overline{\na_{(\m} \na_\n \na_\r \na_\lambda \na_\a \na_{\b)} a_{n-3}} \, , 
\end{align*}
\endgroup
etc., where the overline stands for the coincidence limit.
Define the curvature of the field space:
\be \label{Omega_def}
\left[ \na_\m, \na_\n \right] \varphi^A = \Omega_{\m\n}{}^{A}{}_{B} \varphi^B \, ,
\ee
where $A$, $B$ are matrix-valued indices that represent one Lorentz index for a vector field, a couple of Lorentz indices for the metric perturbations.
Explicitly,
\bear
\Omega^{(\d\tv)}_{\m\n\, ,\,}{}^\a{}_\r &= R_{\m\n}{}^\a{}_\r \, ,
\\
\Omega^{(h)}_{\m\n\, ,\,\a\b}{}^{\r\l} &= 2 R_{\m\n(\a}{}^{(\r} \d^{\l)}_{\b)} \, ,
\\
\Omega^{(\d T)}_{\m\n\, ,\,\a\b\g\;\d\eta\z} &= g_{\a\d} g_{\g\z} R_{\m\n\b\eta} + g_{\a\d} g_{\b\eta} R_{\m\n\g\z} + g_{\g\z} g_{\b\eta} R_{\m\n\a\d} \, ,
\\
\Omega^{(\d\tp)}_{\m\n\, ,\,\a\b\g}{}^{\d\eta\z} &=
\Omega^{(\d T)}_{\m\n\, ,\,\a\b\g}{}^{\d\eta\z} - \Omega^{(\d T)}_{\m\n\, ,\,[\a\b\g]}{}^{\d\eta\z} - \frac{2}{3} g_{\b[\g|}\, \Omega^{(\d T)}_{\m\n\, ,\,\a]}{}^\s{}_\s{}^{\d\eta\z} \, .
\eear
where $\Omega^{(\d T)}_{\m\n}$ has to be additionally antisymmetrized in both $[\a\b\g]$ and $[\d\eta\z]$.
These expressions give, in particular,
\be
\tr\left( \Omega^{(\d\tv)} \right)^2 = - R_{\a\b\g\d}^{\,2} \, ,
\quad\quad
\tr\left( \Omega^{(h)} \right)^2 = -6 R_{\a\b\g\d}^{\,2} \, ,
\quad\quad
\tr\left( \Omega^{(\d\tp)} \right)^2 = -12 R_{\a\b\g\d}^{\,2} \, .
\ee
The heat kernel coefficients for the minimal operator $\D = - \Box + E$ can be expressed as local functionals of the Riemann and the configuration space curvatures \cite{Barvinsky:1985an, Groh:2011dw}:
\begingroup
\allowdisplaybreaks
\begin{align*}
\label{off_diag_heat_kernel_coefficients}
\overline{a_0} &= 1 \, ,\\
\overline{\nabla_\m a_0} &= 0 \, ,\\
\overline{\nabla_{(\m} \nabla_{\n)} a_0} &= \frac16 R_{\m\n} \, ,\\
\overline{\nabla_{(\a} \nabla_\n \nabla_{\m)} a_0} &= \frac14 \nabla_{(\a} R_{\m\n)} \, ,\\ \numberthis
\overline{\nabla_{(\a} \nabla_\b \nabla_\m \nabla_{\n)} a_0} &= 
\frac{1}{15} R_{\g(\a|\d|\b} R^{\g}{}_{\m}{}^{\d}{}_{\n)}
+ \frac{1}{12} R_{(\a\b} R_{\m\n)}
+ \frac{3}{10} \nabla_{(\a} \nabla_\b R_{\m\n)} \, , \\
\overline{a_1} &= \frac{1}{6} R - E \, ,\\
\overline{\nabla_\m a_1} &= \frac{1}{12} \nabla_\m R - \frac{1}{6} \nabla^\n \Omega_{\n\m}
-\frac{1}{2} \nabla_\m E \, ,\\
\overline{\nabla_{(\m} \nabla_{\n)} a_1} &=
\frac{1}{90} R^{\a\b\g}{}_\m R_{\a\b\g\n}
+\frac{1}{90} R_{\a\b} R^\a{}_\m{}^\b{}_\n
-\frac{1}{45} R_{\m\a}R^\a{}_\n 
+\frac{1}{36} R R_{\m\n} 
-\frac{1}{60} \D R_{\m\n} 
\\
&+\frac{1}{20} \nabla_{(\m} \nabla_{\n)} R 
+ \frac16 \Omega_{\a(\m} \Omega^\a{}_{\n)}
- \frac16 \nabla_{(\m} \nabla^\a \Omega_{\a|\n)}
- \frac16 R_{\m\n} E
-\frac13 \nabla_{(\m} \nabla_{\n)} E \, ,\\
\overline{a_2} &= \frac{1}{180} R_{\m\n\a\b}^{\,2} - \frac{1}{180} R_{\m\n}^{\,2} - \frac{1}{30} \D R +\frac{1}{72} R^2 + \frac{1}{12} \Omega_{\m\n}^{\,2} - \frac{1}{6} R E + \frac{1}{2} E^2 + \frac{1}{6} \D E \, .
\end{align*}
\endgroup
Here, the coincidence limit is understood.
Note that in the above formula, the identities $1$, the operators $\Omega^2$, defined in \eqref{Omega_def}, and the endomorphism $E$ should be understood as operators acting in the configuration space. 
We show the explicit expressions for the off-diagonal heat kernel coefficient of the operator $\D = - \Box$ that acts on the hook-antisymmetric torsion for future references:
\begingroup
\allowdisplaybreaks
\begin{align*}
\tr\; \overline{a_0 (\D_\tp)} &= 16 \, ,
\quad\quad\quad\quad
\tr\; \overline{\na_{\m} a_0 (\D_\tp)} = 0 \, ,
\\ \numberthis
\tr\; \overline{\na_{(\m} \na_{\n)} a_0 (\D_\tp)} &= \frac{8}{3} R_{\m\n} \, ,
\quad\quad\quad
\tr\; \overline{\na_{(\m} \na_\n \na_{\r)} a_0 (\D_\tp)} = 4 \na_{(\m} R_{\n\r)} \, ,
\\
\tr\; \overline{\na_{(\m} \na_\n \na_\r \na_{\s)} a_0 (\D_\tp)} &= \frac{1}{270} \Big( 6 R_{\a\b\g\d}^{\;2} + 16 R_{\a\b}^{\,2} + 5 R^2 \Big) \left( g_{\m\n}g_{\r\s} + g_{\m\r}g_{\n\s} + g_{\m\s}g_{\n\r} \right) \, ,
\\
\tr\; \overline{a_1 (\D_\tp)} &= \frac{8}{3} R \, ,
\quad\quad\quad\quad
\tr\; \overline{\na_\m a_1 (\D_\tp)} = \frac{4}{3} \na_\m R \, ,
\\
\tr\; \overline{\na_{(\m} \na_{\n)} a_1 (\D_\tp)} &= \frac{1}{90} \Big( - 41 R_{\a\b\g\d}^{\;2} - 4 R_{\a\b}^{\,2} + 10 R^2 \Big) g_{\m\n} \, ,
\\
\tr\; \overline{a_2 (\D_\tp)} &= \frac{1}{45} \Big( - 41 R_{\a\b\g\d}^{\;2} - 4 R_{\a\b}^{\,2} + 10 R^2 \Big) \, .
\end{align*}
\endgroup
We have abbreviated $R_{\m\n\r\s}^{\;2} = R^{\m\n\r\s} R_{\m\n\r\s}$, $R_{\m\n}^{\,2} = R^{\m\n} R_{\m\n}$ throughout the text.

\end{appendix}

\end{CJK*} 
\makeatletter 
\interlinepenalty=10000 
\bibliographystyle{jhep-mymod}
\bibliography{main}
\makeatother 

\end{document}

%% file: preamble.tex
\usepackage[all]{hypcap} 

\usepackage{amsmath}
\usepackage{tikz}
\usepackage{nccmath}

\newcommand\numberthis{\addtocounter{equation}{1}\tag{\theequation}} 
\newcommand{\mycomment}[1]{}
\usepackage{comment}
\usepackage{setspace}

\usepackage{graphicx}
\usepackage{dcolumn}
\usepackage{bm}
\usepackage{subcaption}

\def\ie{{\it i.e.\;}}

\def\tr{{\rm tr}}
\def\Tr{{\rm Tr}}
\def\Log{{\rm \,log\,}}
\newcommand{\nn}{\nonumber}

\newcommand{\be}{\begin{equation}}
\newcommand{\ee}{\end{equation}}
\newcommand{\bea}{\begin{eqnarray}}
\newcommand{\eea}{\end{eqnarray}}
\newcommand{\bear}{\begin{equation}\begin{aligned}}
\newcommand{\eear}{\end{aligned}\end{equation}}
\newcommand{\bes}{\begin{subequations}\begin{align}}
\newcommand{\ees}{\end{align}\end{subequations}}

\let\a=\alpha \let\b=\beta \let\g=\gamma \let\d=\delta
\let\z=\zeta     \let\l=\lambda
\let\m=\mu \let\n=\nu \let\x=\xi \let\r=\rho
\let\s=\sigma \let\t=\tau   
  
\let\G=\Gamma \let\D=\Delta  \let\L=\Lambda

\def\cG{{\cal G}}

\newcommand{\lctens}{\eta}
\newcommand{\na}{\nabla}
\newcommand{\tv}{T}
\newcommand{\ta}{a}
\newcommand{\tp}{\t}

\newcommand{\polov}{\frac{1}{2}}
\newcommand{\ofbox}{(\Box)}

\let\lamBar=\z  

\newcommand{\logdivsubscript}{{\rm log.div.}}